\documentclass[trackchanges, twocolumn,tighten,twocolappendix]{aastex631}

\immediate\write18{texcount -sub=section -inc main.tex > count.txt}
\usepackage{verbatim}

\defcitealias{jiang2021ApJS}{J21}

\usepackage{CJKutf8}

\usepackage{enumitem}
\setlist{nolistsep}
\setlist[itemize]{align=parleft,left=0pt..1.5em}
\setlist[enumerate]{align=parleft,left=0pt..1.5em}
\usepackage{multirow}
\usepackage{amsmath}

\usepackage{listings}

\newcommand\vectoraas[1]{{\boldsymbol{#1}}}

\shorttitle{AREPO-RSG: Pulsation-lifted CSM}
\shortauthors{Ma et al.}

\begin{document}
\begin{CJK*}{UTF8}{gbsn}


\title{AREPO-RSG: Aspherical Circumstellar Material and Winds from Pulsating Dusty Red Supergiants in Global 3D Radiation Hydrodynamic Simulations}


\author[0000-0002-9911-8767]{Jing-Ze Ma (马竟泽)}
\affiliation{Max Planck Institute for Astrophysics, Karl-Schwarzschild-Str. 1, 85748 Garching, Germany}\email{jingze@mpa-garching.mpg.de}

\author[0000-0001-7969-1569]{Stephen Justham}
\affiliation{Max Planck Institute for Astrophysics, Karl-Schwarzschild-Str. 1, 85748 Garching, Germany}

\author[0000-0003-3308-2420]{Rüdiger Pakmor}
\affiliation{Max Planck Institute for Astrophysics, Karl-Schwarzschild-Str. 1, 85748 Garching, Germany}

\author[0000-0003-3891-7554]{Andrea Chiavassa}
\affiliation{Universit\'e C\^ote d'Azur, Observatoire de la C\^ote d'Azur, CNRS, Lagrange, CS 34229, Nice,  France}
\affiliation{Max Planck Institute for Astrophysics, Karl-Schwarzschild-Str. 1, 85748 Garching, Germany}

\author[0000-0003-2012-5217]{Taeho Ryu}
\affiliation{JILA, University of Colorado and National Institute of Standards and Technology, 440 UCB, Boulder, CO 80308, USA}
\affiliation{Department of Astrophysical and Planetary Sciences, 391 UCB, Boulder, CO 80309, USA}
\affiliation{Max Planck Institute for Astrophysics, Karl-Schwarzschild-Str. 1, 85748 Garching, Germany}

\author[0000-0001-9336-2825]{Selma E. de Mink}
\affiliation{Max Planck Institute for Astrophysics, Karl-Schwarzschild-Str. 1, 85748 Garching, Germany}

\begin{abstract}
Recent observations have revealed a surprisingly large fraction of hydrogen-rich supernovae (SNe) interacting with dense confined circumstellar material (CSM), whose origin is heavily debated. 
Exploiting our recent implementation of a sophisticated radiation transport scheme in the moving-mesh code \texttt{AREPO}, we perform full-sphere 3D radiation hydrodynamic simulations of red supergiant envelopes.
For $10\, M_\odot$ and $20\, M_\odot$ core-carbon-burning stars,
we find that large-amplitude radial pulsations lift the  surface material of density $10^{-14}$--$10^{-12}\; \mathrm{g\; cm^{-3}}$ to the circumstellar environment up to $3\times10^{14}$ cm, consistent with the inferred density for the interacting SN 2013fs.
There, radiation acts on dust to drive highly anisotropic outflows of $10^{-6}$--$10^{-5}\, M_\odot\, \mathrm{yr^{-1}}$.
The total CSM masses for both simulations are $\sim 0.01\, M_\odot$.
Due to convection, the CSM density structure has order-of-magnitude angular variations, dominated by large-scale asymmetries.
We suggest that (1) the CSM around the progenitor is bound material instead of a widely-assumed steady wind, (2) highly aspherical CSM is common and can be created by surface convection rather than only from binary interactions, and (3) 3D effects need to be incorporated in 1D SN modeling, potentially via effective clumping.
Based on our simulations, we propose a 1D analytical CSM model to be directly used for SN observable modeling.
We predict that progenitor pulsations (seen in SN 2023ixf) and highly-confined CSM (seen in SN 2013fs) should be common among most hydrogen-rich SNe.
This can be tested with progenitor monitoring using Rubin Observatory and near-future high-cadence surveys such as ULTRASAT and UVEX.
\end{abstract}


\section{Introduction}
\label{sec:intro}

Modern photometric surveys and spectroscopic observations discovered a significant fraction of supernovae (SNe) showing evidence of interactions (interacting SNe), e.g., enhanced luminosity, delayed shock breakout, or transient narrow emission lines \citep[see e.g.,][for reviews]{smith2017HandbookofSupernovae, dessart2024arXive-prints}.
These interacting SNe indicate that the progenitor stars do not explode in vacuum, but in dense, confined circumstellar material (CSM).
This opens up the possibility to probe the stellar evolution and mass loss at late stages, which were previously not accessible from observations of normal stars.

Observationally, most constraints on the CSM come from hydrogen-rich SNe (Type II), which are core-collapse SNe of evolved massive stars with hydrogen-rich envelopes.
The progenitor stars are predominantly red supergiants (RSGs) and a small population of yellow or blue supergiants and luminous blue variables \citep{smartt2009ARA&A}.
The observed fraction of interacting SNe in the entire Type II SNe population is at least $30\%$--$40\%$ \citep{bruch2021ApJ, bruch2023ApJ, hinds2025MNRAS}, but the actual fraction is likely higher \citep{morozova2018ApJ, forster2018NatureAstronomy}.
The derived CSM density is typically $> 10^{-14}\, \mathrm{g\, cm^{-3}}$ within $10^{14}$--$10^{15}$ cm from the center of the star \citep[or $1.5$--$15$ stellar radii for a $1000\, R_\odot$ RSG; e.g.,][]{yaron2017Nat.Phys., zimmerman2024Nature, jacobson-galan2024ApJa}.
This indicates the presence of dense, confined CSM close to the progenitor stars.
Most analyses assume a wind-like CSM profile to interpret the observational data.
The inferred mass loss rates are $> 10^{-4}\, M_\odot\, \mathrm{yr^{-1}}$ or even reaching $1\, M_\odot\, \mathrm{yr^{-1}}$ \citep[e.g.,][]{fransson2014ApJ, boian2020MNRAS, hinds2025MNRAS, ransome2025ApJ}, at least two orders of magnitude higher than the mass loss rate of core-helium-burning RSGs \citep[e.g.,][]{dejager1988A&AS, beasor2020MNRAS, antoniadis2024A&A}.

These leave us a theoretical puzzle: What is the origin of the dense CSM?
One hypothesis is wave-driven mass loss, as proposed by \citet{quataert2012MNRAS}, where core convection triggers gravity waves that propagate outwards and launch an eruptive wind \citep{shiode2014ApJ, fuller2017MNRAS}.
However, later studies showed that the wave-heating rate is not strong enough to drive significant mass ejections \citep{mcley2014MNRAS, wu2021ApJ, wu2022ApJ, leung2021ApJ}.
Alternatively, the explosive burning may unbind part of the envelope directly without invoking gravity waves, even though this mechanism is limited to a small progenitor mass range \citep{smith2014ApJ, woosley2015ApJ}.
Another hypothesis is mass ejection during binary interactions \citep[e.g.,][]{smith2014ApJ, mcley2014MNRAS, ouchi2017ApJ, matsuoka2024ApJ, ercolino2024A&A}, but whether this channel can explain all of the interacting SNe is not clear:
Although a large fraction of hydrogen-rich SNe are thought to be binary products \citep{zapartas2019A&A, ercolino2025arXive-prints},
only a small fraction of them may interact shortly before the explosion \citep[e.g.,][]{kozyreva2022ApJ, ercolino2024A&A}.

The CSM may also be purely related to the stellar surface variability, in particular for RSGs -- the predominant progenitors of Type II SNe.
RSGs are known to be large-amplitude pulsators \citep{kiss2006MNRAS}, have large-scale surface convection \citep{gilliland1996ApJ}, and are surrounded by extended atmospheres \citep{arroyo-torres2015A&A}.
The Great Dimming of Betelgeuse is an example of possible mass ejections from RSGs \citep{montarges2021Nature, dupree2022ApJ}.

It was suggested that the extended atmosphere of RSGs may be enough to explain the CSM \citep{dessart2017A&A}, but the origin of the extended atmosphere is debated.
Analytical arguments suggest that bound material may be lifted from the surface by convection and pulsation \citep{soker2021ApJ} or shocks generated by transonic convection \citep{fuller2024OJAp}.
Large-amplitude pulsations \citep[][]{goldberg2020ApJ, bronner2025arXive-prints, laplace2025arXive-prints} or convection \citep{goldberg2022ApJa} may change the density structure of the pre-SN RSG, thereby changing the lightcurves.
Using 3D simulations, \citet{goldberg2025arXive-prints} found that yellow supergiants can launch eruptive mass loss via large-amplitude pulsations.
For RSGs, it was also proposed that the pulsation amplitude grows when the luminosity-to-mass ratio ($L/M$) is large, resulting in episodic mass loss \citep{heger1997A&A, yoon2010ApJa, sengupta2025arXive-prints, suzuki2025arXive-prints}, but the actual mass loss process has only been simulated once in 1D simulations \citep{clayton2018} and is difficult to reproduce due to numerical issues (Vincent Bronner priv. comm.).

Such pulsations are already present in 3D simulations of RSGs \citep{chiavassa2024LRCA, goldberg2022ApJ}, but none of the current 1D or 3D models so far successfully produce a CSM dense enough to explain the interacting Type II SNe.
A potential issue is that 1D hydrodynamic models break before mass ejections \citep{bronner2025arXive-prints, suzuki2025arXive-prints} and, as we suggest in this work, 3D models do not include enough envelope for waves to fully steepen into shocks.

Recently, we overcame those technical barriers by demonstrating a RSG simulation an order of magnitude deeper than other previous simulations \citep{ma2025Submitt.AA}.
This is credited to the flexible mesh refinement and local time-stepping supported by our radiation transport module \texttt{AREPO-IDORT} \citep{ma2025Submitt.AA} in the moving-mesh code \texttt{AREPO} \citep{springel2010MNRAS}.
This enables us to perform a radiation hydrodynamic simulation spanning 6 orders of magnitude in time-scale and 4 orders of magnitude in length-scale \citep[see Figure 18 in][]{ma2025Submitt.AA}.
In this work, we present a subset of our 3D \texttt{AREPO-RSG} model grid focusing on the later evolutionary stages, targeting at producing circumstellar material self-consistently from pre-SN RSGs.


\section{Methods}
\label{sec:method}


\begin{figure*}[htb!]
\centering
\includegraphics[width=0.8\textwidth]{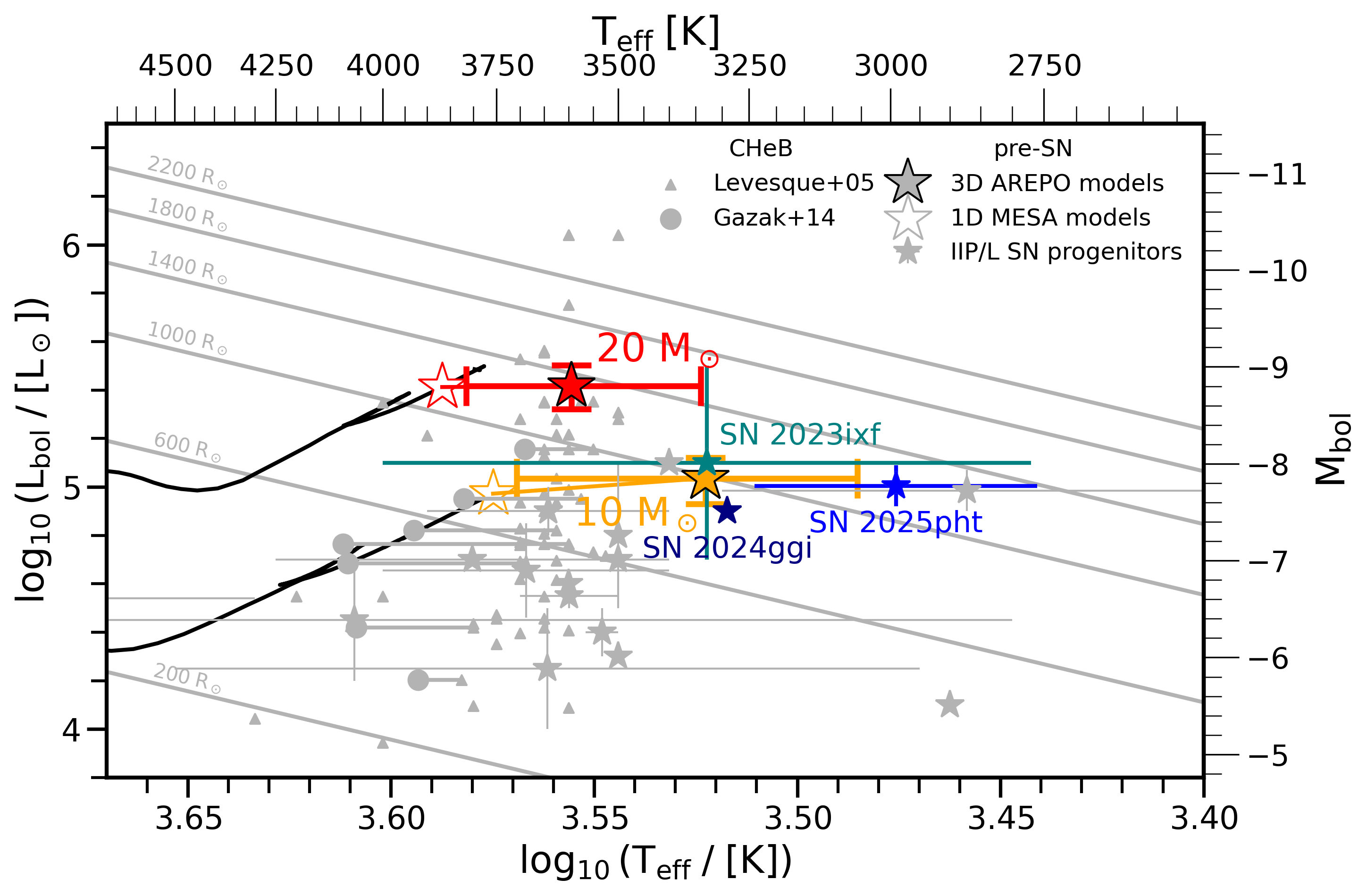}
\caption{
Locations of two 3D pre-SN \texttt{AREPO-RSG} simulations on the Hertzsprung-Russell diagram.
The evolutionary tracks of \texttt{MESA} models are indicated in black solid lines.
We mark the 1D \texttt{MESA} models (empty stars) as initial conditions for the 3D simulations (filled stars with errorbars indicating the $3\sigma$ variations due to temporal variability).
The effective temperatures for the simulations are spherically-averaged (see Appendix~\ref{apx:method:analysis}).
Background gray scatter dots indicate observed Galactic RSG population obtained through TiO bands \citep[triangles;][]{levesque2005ApJ} or SED fitting \citep[circles;][]{gazak2014ApJ}.
The small gray stars mark the Type IIP/L supernova progenitors identified in pre-explosion images compiled by \citet{vandyk2025Galaxies}, where we highlight the well-studied interacting SNe 2023ixf and 2024ggi.
We also highlight the dusty progenitor of SN 2025pht detected by the James Webb Space Telescope \citep{kilpatrick2025arXive-prints}.
Both the absolute values and the uncertainties of the bolometric luminosity may be significantly underestimated \citep{beasor2025ApJ}.
Background gray lines show contours of constant radius.
\label{fig:0d:hr}}
\end{figure*}

We use the 3D finite-volume moving-mesh code \texttt{AREPO} \citep{springel2010MNRAS, pakmor2016MNRAS, weinberger2020ApJS} to perform two radiation hydrodynamic simulations of core-carbon-burning RSG envelopes.
A detailed description of 1D initial conditions and numerical methods are presented in Appendix~\ref{apx:method}, where the stellar parameters and numerical parameters are summarized in Table~\ref{tab:sim}.
The limitations of our simulations are further discussed in Appendix~\ref{apx:discussion:caveat}.
Here, we only provide a brief overview.

First, we use the 1D stellar evolution code \texttt{MESA} \citep[version 15140; ][]{paxton2011ApJS, paxton2013ApJS, paxton2015ApJS, paxton2018ApJS, paxton2019ApJS, jermyn2023ApJS} to construct 1D non-rotating RSG profiles at near-solar metallicity ($Z=0.02$).
We select a $10\, M_\odot$ RSG (initially $11.5\; M_\odot$) at $200$ years before core collapse and a $20\, M_\odot$ RSG (initially $22\; M_\odot$) at $8000$ years before core collapse.
We define their radii as $R_\mathrm{MESA}$.
Then, we follow \citet{ohlmann2017A&A} to replace the inner $3\%\, R_\mathrm{MESA}$ with a less dense artificial core in hydrostatic equilibrium, map the modified profile onto \texttt{AREPO}, and damp the velocities globally for one stellar sound-crossing timescale to aid relaxation.
Within the artificial core, we further apply a constant luminosity source and a continuous damping term that we keep after relaxation.
The simulation box is $300\, R_\mathrm{MESA}$ wide, filled with low density $\rho_\mathrm{bg}=10^{-17}\, \mathrm{g\, cm^{-3}}$ and low temperature $T_\mathrm{bg}=530$ K pseudo-vacuum.

The equations of hydrodynamics are solved using a second-order accurate finite-volume approach with an HLLD Riemann solver \citep{springel2010MNRAS, pakmor2016MNRAS}.
We solve full self-gravity using a Oct-tree multiple expansion method \citep{weinberger2020ApJS}.

The radiation transport is coupled to the hydrodynamics via our newly-implemented \texttt{AREPO-IDORT} module, which solves for gray specific intensities at discrete directions by solving the time-independent radiative transfer equations using an implicit first-order discrete ordinates method \citep{ma2025Submitt.AA}.
The radiation transport is performed on the globally synchronized hydrodynamic timesteps.

We include a realistic equation of state and gray opacity tables in our simulations.
We use the high-temperature \texttt{OPAL} equation of state \citep{rogers2002ApJ} blended with ideal gas below a temperature of 1870 K.
We use Rosseland and Planck opacities from the high-temperature \texttt{Los Alamos OPLIB} table \citep{colgan2016ApJ}\footnote{\url{https://aphysics2.lanl.gov/apps/}  The opacity due to electron scattering is included in the Rosseland opacity table.} stitched to a low-temperature \citet{ferguson2005ApJ} table below a temperature of 30000 K.\footnote{\url{https://www.wichita.edu/academics/fairmount\_las/physics/Research/opacity.php}}
These EOS and opacity tables cover the temperature and density range of RSG envelopes and include the ionization/recombination between ionized species and atomic species.
The low-temperature opacity table also includes contributions from molecular lines and dust in equilibrium chemistry \citep{ferguson2005ApJ}.
We use the Rosseland opacity as the flux-weighted opacity, and use the Planck opacity as the energy-weighted opacity.

We do not explicitly model dust in our simulations, but we include the effects of radiation acting on dust through opacity.
Below $1200$~K in the optically-thin regions, we take the Planck opacity values as both the energy-weighted and the flux-weighted opacities to take into account the high opacities from dust.
Between $1500$--$1200$~K, we take a linear interpolation between the Rosseland opacity and the Planck opacity for the opacity value to guarantee a smooth transition.
We therefore make the implicit assumption that the dust forms instantly in chemical equilibrium with other species according to \citet{ferguson2005ApJ}, with the size distribution from \citet{mathis1977ApJ}.

Each simulation consists of  approximately $20$ million cells and took 3.5 months to reach 70 stellar years running on 504 CPU cores (1.3 million CPU hours).
The radiation transport \texttt{AREPO-IDORT} module takes about half of the computational cost.
Given the flexible mesh construction in \texttt{AREPO}, we employ a multi-shell refinement criterion, where we set different target volume resolutions at different radii (see Appendix~\ref{apx:method:refinement}).
The Voronoi mesh is allowed to move with the gas in a quasi-Lagrangian way, which gives us the advantage of resolving the outflow.

Throughout this paper, we only present analyses of the simulations after they reach steady states.
We determine the steady state as when the mass ejection is not influenced by the initial transient ejection anymore, i.e. the spherically-averaged density contour at $10^{-16}\; \mathrm{g\; cm^{-3}}$ begins to rise again with time.
But we also check that two extra conditions are met during the steady state: (1) Both the bolometric luminosity and spherically-averaged radius vary around an approximately constant value, as in 3D \texttt{CO$^5$BOLD} simulations \citep{ahmad2023A&A}; (2) The temporally-averaged total radial energy flux is approximately spatially constant, as in 3D \texttt{Athena++} simulations \citep{goldberg2022ApJ, goldberg2025arXive-prints}.
Detailed checks are presented in Appendix~\ref{apx:method:steadystate}.
How we obtain the spherically-averaged and global quantities, e.g., luminosity, radius and effective temperature, is presented in Appendix~\ref{apx:method:analysis}.

\section{Results}

We show the locations of our two 3D simulations on the Hertzsprung-Russell diagram in Figure~\ref{fig:0d:hr}.
They appear redder than the 1D models due to more extended superadiabatic layers (Appendix~\ref{apx:disc:radius}), but are consistent with the derived luminosities and effective temperatures of observed Type IIP SN progenitors.

\subsection{Episodically-lifted CSM via Radial Pulsation}

\begin{figure*}[htb!]
\centering
\includegraphics[width=\textwidth]{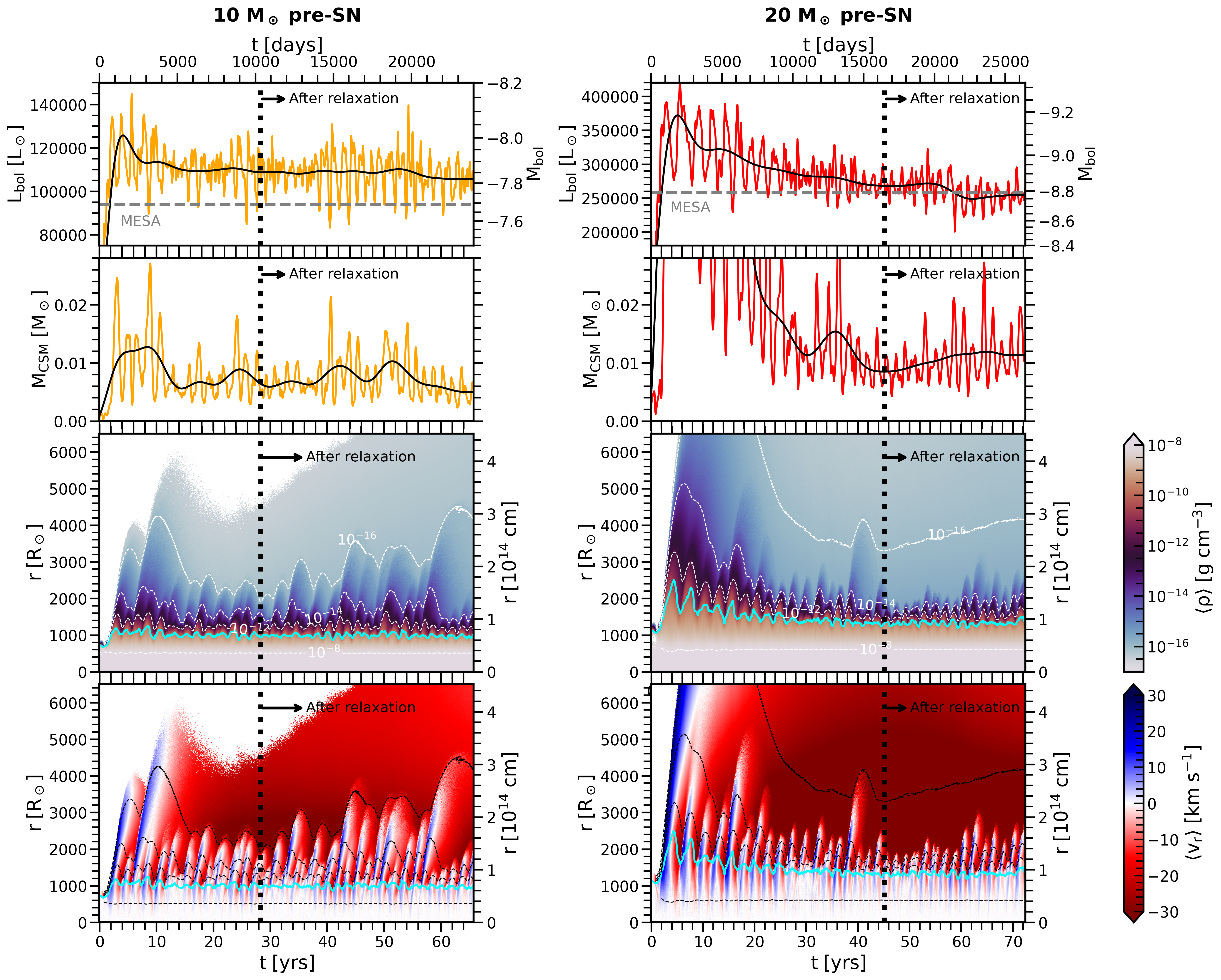}
\caption{
Dense CSM episodically lifted by semi-regular pulsation in the $10\, M_\odot$ simulation (left) and $20\, M_\odot$ simulation (right).
From top to bottom, we show the bolometric luminosity, CSM mass, spherically-averaged density, and spherically-averaged radial velocity.
In the top two rows, black solid lines show the general trends of variations filtering out the high-frequency variability.
The bolometric luminosity from the \texttt{MESA} models used as initial conditions are indicated with gray dashed horizontal lines.
In the bottom two rows, contour lines indicate the iso-density levels $[10^{-8},10^{-12},10^{-14},10^{-16}]$ $\mathrm{g\, cm^{-3}}$.
The cyan lines indicate the spherically-averaged Rosseland radius defined in Appendix~\ref{apx:method:analysis}.
The right-hand side of the vertical dotted lines indicate the relaxed phase as defined at the end of Section~\ref{sec:method}.
\label{fig:0d:tvar}}
\end{figure*}



After the initial relaxation phase, we find that both simulations settle into steady states with semi-regular variability in bolometric luminosity. The magnitude of this time variability is represented by the errorbars for the 3D simulations in Figure~\ref{fig:0d:hr}, and shown in the top row of Figure~\ref{fig:0d:tvar}.
The dominant periods align with expectations of the fundamental modes for RSGs (see Appendix~\ref{apx:analyses:fourier}).
The mechanism that sustains the radial pulsations is likely the $\kappa\gamma$ mechanism, where the opacity peak due to hydrogen and helium recombination leads to unstable growth of perturbations \citep{heger1997A&A, joyce2020ApJ} with a non-linear boost from recombination energy \citep{clayton2018, bronner2025arXive-prints}.
We defer detailed analyses of pulsation properties to future explorations.

The large-amplitude radial pulsation acts as a piston that pushes the dense material from the stellar surface into the circumstellar environment.
As shown in the second row of Figure~\ref{fig:0d:tvar}, the lifted material forms CSM of $\sim 0.01\; M_\odot$ in both simulations.
We define the CSM mass as the total mass of material with density between $10^{-16}$--$10^{-10}\; \mathrm{g\; cm^{-3}}$ to exclude the background pseudo-vacuum and the stellar interior.

In the bottom two rows of Figure~\ref{fig:0d:tvar}, we plot the spherically-averaged density $\langle\rho\rangle$ and radial velocity $\langle v_r\rangle$.
As shown in the radial velocity inside the stars, the dominant internal motion is the global contraction and expansion of the entire envelope, which further supports that the radial pulsation is governed by the fundamental mode.
The large-amplitude pulsations lift the dense material from the stellar surface.
The lifted material is then slowed down by both gravitational pull and collision with the pre-lifted material.
Eventually, most of the material falls back and collides with the material lifted in the next several pulsations.
Over time, this develops into a quasi-static CSM structure where the inner CSM close to the stellar surface varies at the pulsation period, while the outer CSM at $2\times 10^{14}$ cm varies at a longer timescale several times larger than the pulsation period.

\subsection{Spherically-averaged Density Profiles: A Two-zone Model}
\label{sec:result:agb}


\begin{figure*}[htb!]
\centering
\includegraphics[width=0.8\textwidth]{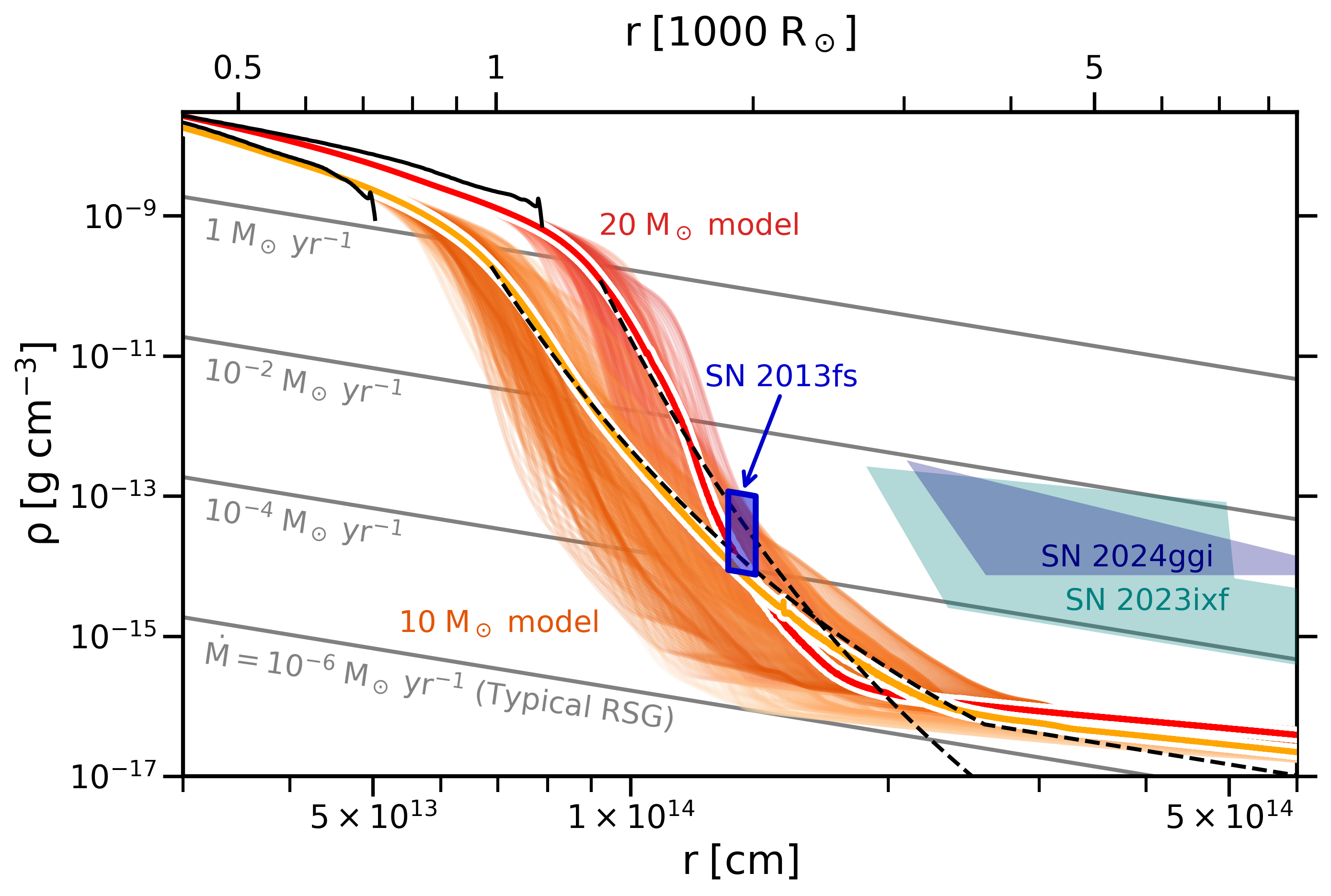}
\caption{
The CSM density structure from our 3D simulations is consistent with the range of densities inferred for SN 2013fs.
Different colored lines indicate the spherically-averaged density profiles at different times, for the $10\, M_\odot$ simulation (orange) and the $20\, M_\odot$ simulation (red), which are much more extended than the initial conditions from \texttt{MESA} (black solid lines).
The white-edged solid lines represent the density profiles averaged both in time and angles.
The black dashed lines show the density profiles described by the analytical `two-zone model' detailed in Section~\ref{sec:result:agb}.
We also highlight the inferred density profile from three well-studied interacting SNe, SN 2013fs \citep{yaron2017Nat.Phys.}, 2023ixf \citep[compiled by][]{nayana2025ApJ} and SN 2024ggi \citep{zhang2024ApJa, jacobson-galan2024ApJ, shrestha2024ApJ, ertini2025A&A, chen2025ApJ}, as labeled.
The background gray solid lines indicate density profiles for constant mass-loss rates assuming a constant wind velocity $30\, \mathrm{km\, s^{-1}}$.
\label{fig:1d:rhor}}
\end{figure*}


In Figure~\ref{fig:1d:rhor}, we show the spherically-averaged CSM density profiles of the $10\, M_\odot$ (orange) and $20\, M_\odot$ simulation (red).
Different curves indicate the spherically-averaged density profiles at different times.
The white-edged curves indicate the spherically-and-temporally-averaged density profiles.
Deep inside the stars, the density structures do not vary significantly with time, and agree well with the initial profiles from \texttt{MESA} (black solid lines).
For reference, in gray solid lines, we also plot the density structure from steady winds assuming a constant wind velocity $30\, \mathrm{km\, s^{-1}}$ for mass-loss rates of $10^{-6}$--$1\, M_\odot\, \mathrm{yr^{-1}}$.

Our simulations predict a two-zone CSM density structure: a dense quasi-static CSM confined within $3\times 10^{14}$ cm attached to a less dense CSM due to a dust-driven wind outside.
Such a two-zone CSM structure was also found necessary to explain some interacting SNe, e.g., SN 2013fs \citep{yaron2017Nat.Phys.}, SN 2020tlf \citep{jacobson-galan2022ApJ}, SN 2023ixf \citep{singh2024ApJ, zimmerman2024Nature, nayana2025ApJ}, SN 2024ggi \citep{ertini2025A&A}, and PS1-11aop \citep{ibik2025ApJ}.

We find that the CSM density values in our simulations are consistent with the CSM density inferred for SN 2013fs \citep{yaron2017Nat.Phys.}, and approximately one order of magnitude less than SN 2023ixf \citep[compiled by][]{nayana2025ApJ} and SN 2024ggi \citep{zhang2024ApJa, jacobson-galan2024ApJ, shrestha2024ApJ, ertini2025A&A, chen2025ApJ}.
However, there are uncertainties in both our treatment of dust and in deriving the density structure from SN observations.
We expect that forward modeling from our simulation and direct comparison with the observed SN lightcurves and spectra will be a more reliable test of our simulation predictions.

Here, we show that the first dense inner zone of the CSM in our simulation is bound material, as opposed to enhanced mass loss as interpreted in most studies.
Such a dense atmosphere is supported by a train of periodic shocks generated by large-amplitude pulsations (see Figure~\ref{fig:0d:tvar}), as also proposed for the atmospheric structure of asymptotic giant branch (AGB) stars \citep{bertschinger1985ApJ, bowen1988ApJ}.
The density profile as a function of radial distance $r$ can be well-described by a shock-supported quasi-static atmosphere \citep[Equation 5 in][]{fuller2024OJAp}:
\begin{equation}
    \rho(r) = \rho(R)\left(\frac{R}{r}\right)^2 e^{-\frac{v_\mathrm{esc}^2(R)}{2v_\mathrm{s}^2}\left(1-\frac{R}{r}\right)}\, .
\label{eq:rho:static}
\end{equation}
We take $R$ and $\rho(R)$ to be the Rosseland radius and the associated density averaged over time and spherical shells in our simulation.
These two quantities can also be provided by 1D stellar evolution models if a proper convective efficiency is chosen to match the predicted radius in our 3D simulations.
For the shock velocity, we take the sound speed predicted by the 1D \texttt{MESA} model at the location where the acoustic timescale becomes comparable to the radiative cooling timescale, i.e. $\tau=c/c_s$ \citep{jiang2015ApJ, cheng2024ApJ}.
Here, $\tau$ is the optical depth integrated from the stellar surface, $c$ is the speed of light, and $c_s$ is the local sound speed.
This is motivated by the theoretical consideration that the weak shock speed approximately follows the local sound speed until radiative cooling yields a nearly isothermal atmosphere.
For both models, the shock speeds are approximately $11\, \mathrm{km\, s^{-1}}$ (supersonic near the surface), which are consistent with the spherically-averaged radial velocities shown in Figure~\ref{fig:0d:tvar}.
The density profiles in Equation~\ref{eq:rho:static} are shown in Figure~\ref{fig:1d:rhor} as the inner parts of the black dashed lines.

\begin{figure*}[htb!]
\centering
\includegraphics[width=\textwidth]{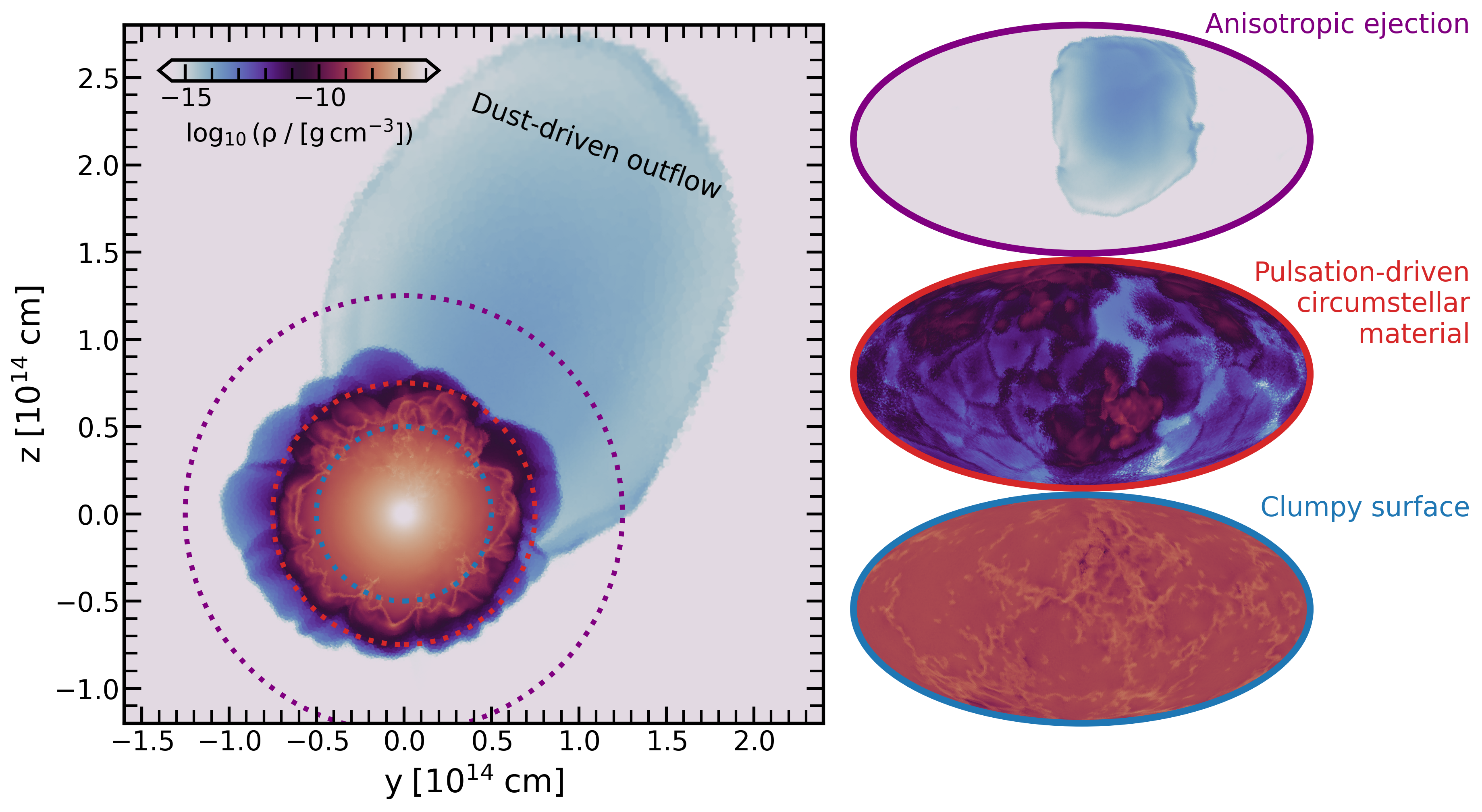}
\caption{
Convection creates a clumpy stellar surface, and aspherical circumstellar material, leading to an anisotropic outflow.
Left: A 2D density slice of the $10\, M_\odot$ simulation at 60.5 years.
Right: Projected density maps along three spheres indicated by the dashed circles in the left plot (where the color at the edge of each density map corresponds to that of the appropriate circle).
An interactive 3D visualization can be found here:
\href{https://jingzema.com/AREPO-RSG/arepo_rsg_csm_fast.html}{https://jingzema.com/AREPO-RSG/arepo\_rsg\_csm\_fast.html}
\label{fig:2d:overview}}
\end{figure*}

The outer zone of the CSM in our simulation is an extended tail due to the dust-driven wind.
As the shock-supported atmosphere extend to large distances, the temperature drops below $1200$--$1500$ K, where dust forms \citep{hofner2018A&AR}, and the radiation force acts on the dust opacities to drive a wind \citep{fuller2024OJAp}.
In our simulations, the wind is driven by radiation acting on the high Planck opacity taken from the \citet{ferguson2005ApJ} table below $1200$--$1500$ K.
We find that the outward radiation force on this material marginally exceeds the attraction from gravity in our simulations, even though the simulations are not long enough for us to see the wind material becoming unbound.
The density profile can also be predicted from analytical theory.
The dust-forming radius can be approximated by $R_\mathrm{d}=(T(R)/T_\mathrm{d})^2 R$ \citep{hofner2018A&AR, fuller2024OJAp}, where $T_\mathrm{d}$ is the dust-forming temperature and the term $T(R)^2 R$ can be derived from luminosity of the \texttt{MESA} model $L=4\pi R^2\sigma T(R)^4$.
Here, $\sigma$ is the Stefan-Boltzmann constant.
The density $\rho(R_\mathrm{d})$ at the dust-forming radius can be found by plugging $R_\mathrm{d}$ in Equation~\ref{eq:rho:static}.
Then the wind mass-loss rate is $\dot{M}=4\pi R_\mathrm{d}^2\rho(R_\mathrm{d})v_\mathrm{s}$, and the density structure for the dust-driven wind is
\begin{equation}
\begin{split}
    \rho(r) & = \frac{\dot{M}}{4\pi r^2 v_\infty} = \rho(R_\mathrm{d})\frac{R_\mathrm{d}^2v_\mathrm{s}}{r^2v_\infty} \\
    & = \rho(R)\left(\frac{R}{r}\right)^2 \frac{v_\mathrm{s}}{v_\infty}\exp\left[-\frac{v_\mathrm{esc}^2(R)}{2v_\mathrm{s}^2}\left(1-\frac{R}{R_\mathrm{d}}\right)\right]\, ,
\label{eq:rho:wind}
\end{split}
\end{equation}
where we assume a terminal wind speed of $v_\infty = 30\, \mathrm{km\, s^{-1}}$ \citep{mauron2011A&A}.\footnote{The order of magnitude of the wind mass-loss rate (and therefore the density) is not sensitive to the terminal wind speed, because the terminal speed mostly ranges from $10$ to $50\, \mathrm{km\, s^{-1}}$ \citep{mauron2011A&A, decin2024A&A}.
The terminal wind speed can also be predicted analytically based on an estimation of the dust opacity \citep[e.g., Equation 14 in][]{fuller2024OJAp}.}
The entire CSM density structures as plotted in black dashed lines in Figure~\ref{fig:1d:rhor} are found by attaching Equation~\ref{eq:rho:static} to this wind solution.
We find that taking $T_\mathrm{d} = 1300$ K for the $10\, M_\odot$ simulation yields  a good fit, while the $20\, M_\odot$ simulation does not drive a wind, potentially due to its small radius and low luminosity for its mass.
The $20\, M_\odot$ star will become more luminous in the final thousand years than simulated here, and therefore is expected to have more violent mass ejections than in our simulation prior to core collapse.
Both the 3D simulations and the analytical descriptions predict a mass loss rate between $10^{-6}$--$10^{-5}\, M_\odot\, \mathrm{yr^{-1}}$, which is broadly consistent with the observed mass-loss rates \citep[e.g.,][]{dejager1988A&AS, beasor2020MNRAS, antoniadis2024A&A, decin2024A&A}.

\subsection{3D Effects Due to Convection: Clumpy Surfaces, Aspherical CSM, and Anisotropic Outflows}

\begin{figure*}[htb!]
\centering
\includegraphics[width=\textwidth]{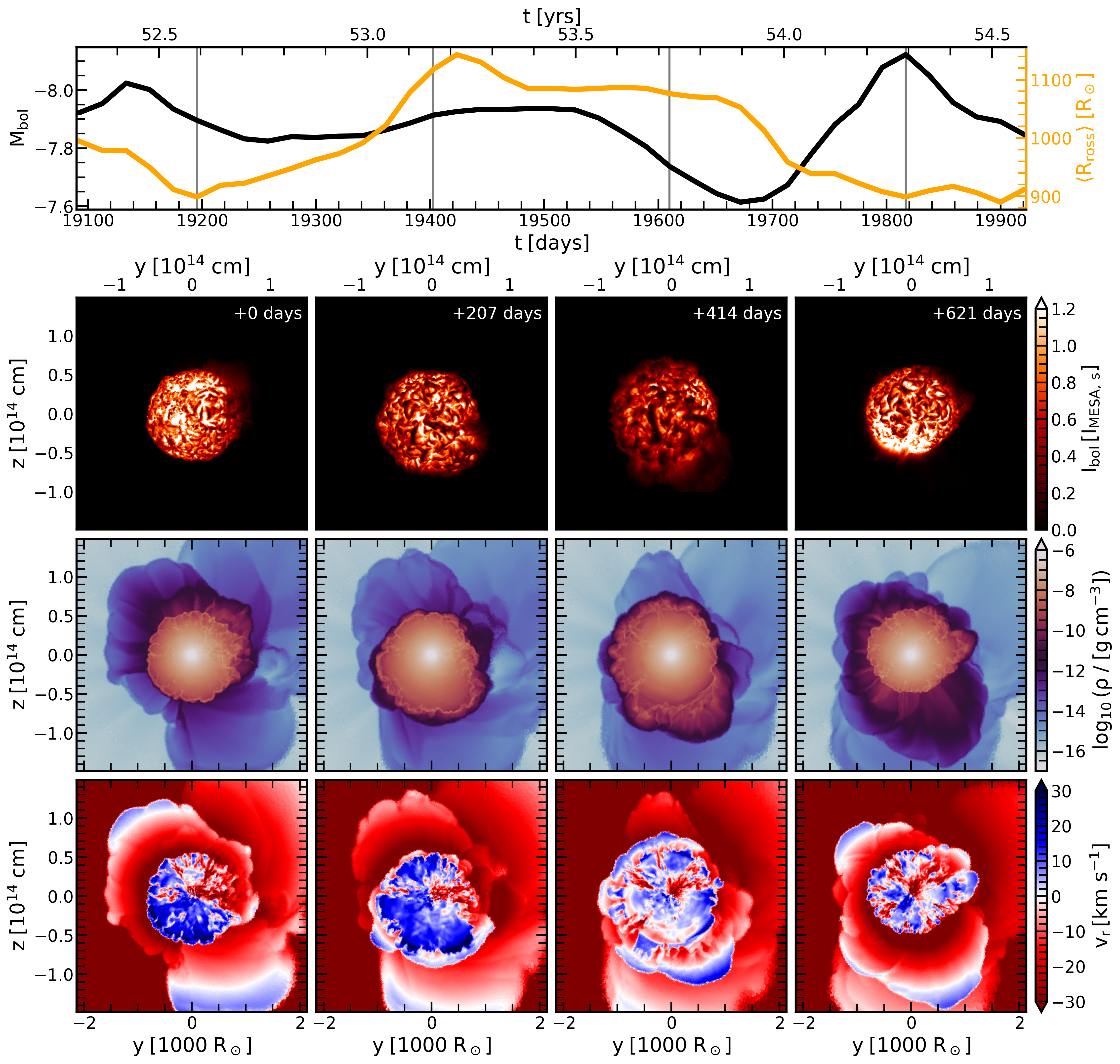}
\caption{
Time sequence of the convective envelope variations within one pulsation cycle for the $10\, M_\odot$ simulation.
The top row shows the variability of the absolute bolometric magnitude $M_\mathrm{bol}$ in black and the spherically-averaged Rosseland radius $\langle R_\mathrm{ross}\rangle$ in orange.
During this one pulsation cycle, we select 4 snapshots equally spaced in time (indicated by gray vertical lines in the top row), and plot the bolometric intensity looking from the x axis (second row), y-z mid-plane slice of the density (third row), and y-z mid-plane slice of the radial velocity (last row).
\label{fig:2d:conv}}
\end{figure*}


In our simulations, large-scale convection in the RSG envelope leads to significant deviations from spherical symmetry, which creates the clumpy surface, aspherical CSM, and anisotropic dust-driven outflows, as illustrated in the left panel of Figure~\ref{fig:2d:overview}.
We take three spheres at different radii (dotted circles in the left panel of Figure~\ref{fig:2d:overview}), and plot the projected density maps along those spheres on the right panels.
Inside the star near the photosphere (bottom right), the density shows small-scale clumps.
The clumpy surface is due to the surface convection driven by radiative cooling near the photosphere (Ma et al. to be subm.).
This is where the supernova shock will propagate through and break out from the surface.
It has been suggested that this clumpy progenitor surface can prolong the shock breakout duration \citep{goldberg2022ApJa}.
In the circumstellar environment (middle right panel in Figure~\ref{fig:2d:overview}), the density fluctuation can still differ by several orders of magnitude in different angles.
This is particularly evident in the ejected material (top right panel in Figure~\ref{fig:2d:overview}), where
the density distribution is dominated by large-scale asymmetries, which reflects the large convective cells inside the star from which the CSM is lifted.

The temporal variations of the convective structure and aspherical circumstellar material are illustrated in Figure~\ref{fig:2d:conv}.
The plot shows the variations during one pulsation cycle for the $10\, M_\odot$ model, for the bolometric intensity (second row), mid-plane slice of density (third row), and mid-plane slice of radial velocity (last row).
Generally, the stellar surface exhibits filament-like structures and dark clumps in the intensity map, almost identical to those seen in the \texttt{CO$^5$BOLD} 3D simulations of AGB stars and RSGs \citep[e.g.,][]{freytag2024A&A}.
The dark clumps are pulsation-lifted dense opaque material shaped by large-scale convection \citep{freytag2024A&A}.
The overall convective pattern on the surface is controlled by the intense cooling, which creates a density inversion subject to Rayleigh-Taylor instability, resulting in dense and low-entropy material mixed into the envelope through fast downdrafts (Ma et al. to be subm.).

On the top row of Figure~\ref{fig:2d:conv}, we show the nearly anti-phase variability between absolute bolometric amplitude $M_\mathrm{bol}$ and spherically-averaged Rosseland radius.
At the luminous phase at 19196 days (defined as $+0$ days in the second row of the plot), the star begins its expansion.
At day $+207$, the expanding star drives shocks and lifts dense material from the surface.
By day $+414$, the star begins to contract.
Most of the dense material falls back onto the star while the outer ejecta expands.
At day $+621$, the star resumes its peak luminosity and begins its next pulsation cycle.
Long filaments appear in the density slice where gas is compressed by shock fronts due to the collision between laterally-expanding ejecta.
Those long filaments are in fact sheets in 3D, and their separation reflects the length scale of deep convection.
This is because the horizontal motions are governed by deep convection (Ma et al. to be subm.), and when a pulse passes near the surface, it steepens into a laterally expanding shock front that lifts the material.
Two adjacent shock fronts will meet at a plane extending radially outwards from the star, which creates the over-dense sheets approximately perpendicular to the stellar surface.
Through multiple pulsation cycles, the star episodically fills its surroundings with dense aspherical material.

We therefore suggest that the SN progenitor profile is clumpy from the stellar surface to the CSM, which should be taken into account in 1D SN modeling, possibly through micro/macroclumping \citep{dessart2018A&A, dessart2019A&Aa}.
This may help explain the different density derived from different wavelengths for interacting SNe \citep{berger2023ApJ, nayana2025ApJ}.
Another naive expectation is that clumpiness allows the radiation to leak out from the low-density chimneys, thereby enhancing the cooling, but detailed radiation hydrodynamic models are needed to access the effects of clumping in progenitors and their associated CSM.

There is a variety of observational evidence that suggests the CSM to be aspherical or clumpy, as indicated by different densities inferred from different wavelengths, multi-peaked emission lines, and polarization signals \citep[e.g.,][]{chandra2012ApJ, smith2015MNRASa, andrews2018MNRAS, andrews2019ApJ, brennan2022MNRAS, kozyreva2022ApJ, smith2023ApJ, vasylyev2023ApJ, bilinski2024MNRAS, singh2024ApJ, shrestha2025ApJ, andrews2025ApJ, nayana2025ApJ, vasylyev2025arXive-prints}.
The aspherical CSM is mostly assumed to be associated with binary interactions \citep[e.g.,][]{smith2015MNRASa, andrews2018MNRAS, brennan2022MNRAS, smith2023ApJ, vasylyev2023ApJ, bilinski2024MNRAS, singh2024ApJ, andrews2025ApJ}.

Here, we show instead that convection can also result in highly aspherical CSM dominated by large-scale structures.
Future spectropolarimetric forward modeling \citep[e.g.,][]{dessart2025A&Aa} from our simulations will be useful to directly compare with observations.

\section{Discussion and Conclusions}
\label{sec:conc}

This is the first of a series of papers where we describe the scientific results from the 3D \texttt{AREPO-RSG} models.
In this work, we perform global 3D radiation hydrodynamic simulations of two pre-explosion RSGs during core carbon burning:
a $10\, M_\odot$ RSG at 200 years before it explodes, and a $20\, M_\odot$ RSG at 8000 years before it reaches core collapse.
Our multi-scale simulations include $97\%$ of the convective envelope in radius and the atmosphere up to $300$ stellar radii.
The differences between our results and other 1D or 3D simulations are discussed in Appendix~\ref{apx:disc} along with our simulation caveats.

We find that dense confined CSM of $\sim 0.01\; M_\odot$ is self-consistently produced in the simulations, which is episodically-lifted by large-amplitude radial pulsations.
We interpret those pulsations as fundamental modes excited by the $\kappa\gamma$-mechanism \citep[e.g.][]{bronner2025arXive-prints}.
The pulsations steepen into shocks and lift the dense surface material to the circumstellar environment up to $3\times10^{14}$ cm, where the dust forms and radiation acts on dust to drive outflows of $10^{-6}$--$10^{-5}\, M_\odot\, \mathrm{yr^{-1}}$.
This process is very similar to the pulsation-enhanced dust-driven wind in AGB stars \citep{hofner2018A&AR}.

The CSM density values from our simulations fit well with the CSM density inferred for SN 2013fs and about one order of magnitude lower than the CSM inferred for SN 2023ixf and SN 2024ggi.
This is already a reasonable agreement considering the uncertainties in both simulations and observational inference.
Based on our simulations, we propose a 1D analytical two-zone model to describe the CSM density profile in Section~\ref{sec:result:agb}.

In our simulations, the CSM and the dust-driven outflow are highly aspherical, dominated by large-scale asymmetries.
This is because the large-scale convection in RSG envelope breaks the spherical symmetry in the material ejection.

We therefore propose that:
\begin{itemize}

    \item[$\bullet$] The confined CSM observed in interacting hydrogen-rich SNe may be bound material episodically-lifted from the surface with a velocity dispersion of $<30\, \mathrm{km\, s^{-1}}$, rather than CSM from mass loss as assumed in most works.
    
    \item[$\bullet$] Highly aspherical CSM -- as inferred from spectroscopy and spectropolarimetry -- can also come from surface convection of single stars, rather than only from binary interactions. If true, most CSM in Type II SNe should be aspherical.
    
    \item[$\bullet$] 3D effects -- including consequences of clumpy stellar surfaces, aspherical CSM, and anisotropic outflows -- should be considered in 1D SN modeling, potentially as effective clumping.

\end{itemize}

Our 3D simulations can be readily used for 1D and 3D simulations of SN explosions, wind launching, and binary interactions.
By taking the 1D slice of a 3D profile in different angles and performing 1D radiation hydrodynamic simulations of SN explosions, we can start to study how the 3D geometry affects the observed lightcurve and spectra evolution.
In addition, the dust-driven outflow observed in our simulations is preliminary and need to be studied further with more detailed physics.
Finally, a subset of these pre-explosion RSGs are likely to be in interacting binaries \citep{ercolino2024A&A}, which may create very extended CSM \citep[e.g.,][]{landri2024MNRAS} or trigger precursors \citep[e.g.,][]{tsuna2024ApJ, tsuna2025ApJ}.

The highly aspherical CSM and mass ejection simulated in this work are quantitatively consistent with observed high-luminosity RSGs or hypergiants, e.g., \texttt{IRC+10420} \citep{humphreys1997ApJ}, \texttt{VY CMa} \citep{smith2001AJ, singh2023ApJ}, \texttt{NML Cyg} \citep{schuster2006AJ, debeck2025A&A}, \texttt{VX Sgr} \citep{chiavassa2022A&Aa}, \texttt{WOH G64} \citep{ohnaka2024A&A, munoz-sanchez2024arXive-prints}, and \texttt{DFK 52} \citep{siebert2025A&A}.
Long-term monitoring and spatially-resolved interferometric observations of them will help reveal their evolutionary stages and their connections to interacting SNe.

We also provide clear predictions to be tested in observations:
For most hydrogen-rich SNe with a final progenitor mass $>10\, M_\odot$, SN 2023ixf-like progenitor pulsation and SN 2013fs-like highly-confined CSM should be present.
Observations of those phenomena are still scarce, limited by our detection capability \citep{dessart2024arXive-prints, vandyk2025Galaxies}.
However, within several years in the future, variable SN progenitors are expected to be more frequently detected in the Legacy Survey of Space and Time (LSST) at Vera C. Rubin Observatory \citep{ivezic2019ApJ, hambleton2023PASP}.
With a wide-field near-ultraviolet (NUV) photometric survey such as ULTRASAT \citep{shvartzvald2024ApJ} in collaboration with early follow-up UV spectroscopy such as UVEX \citep{kulkarni2021arXive-prints} and other ground-based instruments, we will have a chance to detect more SN 2013fs-like early interaction events in the coming several years.

\begin{acknowledgments}
We thank Jim Fuller for providing valuable comments to the early draft.
We thank Sebastian Ohlmann for sharing the module to construct stable 3D AREPO giant stars from MESA profiles.
We thank Mike Lau, Jared Goldberg, Luc Dessart, Fabian Schneider, Vincent Bronner, Philipp Podsiadlowski, Andrei Beloborodov, and Raffaella Margutti for helpful discussions.
A.C. acknowledges support from the French National Research Agency (ANR) funded project PEPPER (ANR-20-CE31-0002).
This research project was partly conducted using computational resources (and/or scientific computing services) at the Max-Planck Computing \& Data Facility. 
The authors gratefully acknowledge the scientific support and HPC resources provided by the Erlangen National High Performance Computing Center (NHR@FAU) of the Friedrich-Alexander-Universität Erlangen-Nürnberg (FAU) under the NHR project b234dd.
NHR funding is provided by federal and Bavarian state authorities.
NHR@FAU hardware is partially funded by the German Research Foundation (DFG) – 440719683.
Part of this work was done when the author was attending the TDE24 program at Kavli Institute for Theoretical Physics (KITP), which is supported in part by grant NSF PHY-2309135.
\end{acknowledgments}

\vspace{5mm}

\software{
\texttt{AREPO} \citep{springel2010MNRAS, pakmor2016MNRAS, weinberger2020ApJS}, 
\texttt{MESA} \citep{paxton2011ApJS, paxton2013ApJS, paxton2015ApJS, paxton2018ApJS, paxton2019ApJS, jermyn2023ApJS},
\texttt{Astropy} \citep{astropycollaboration2013A&A, astropycollaboration2018AJ, astropycollaboration2022ApJ}, 
\texttt{NumPy} \citep{harris2020Nature}, 
\texttt{SciPy} \citep{virtanen2020NatureMethods}, 
\texttt{Matplotlib} \citep{hunter2007Comput.Sci.Eng.}, 
\texttt{Jupyter} \citep{kluyver2016PositioningandPowerinAcademicPublishing:PlayersAgentsandAgendas}
}

\appendix

\section{Detailed Methods}
\label{apx:method}

\begin{table*}[htb!]
\begin{center}
 \caption{Parameters of two 3D pre-SN \texttt{AREPO-RSG} simulations.
 The stellar parameters are listed as the mass $M_\mathrm{tot}$, bolometric luminosity $L_\mathrm{bol}$, radius $R_\mathrm{ross}$ where the Rosseland optical depth $\approx 1$, and effective temperature $T_\mathrm{eff}$.
 All these surface quantities in simulation outputs are averaged over spherical shells as described in Appendix~\ref{apx:method:analysis}, with errorbars indicating the $3\sigma$ variations due to temporal variability.
 The stellar Rosseland radii from \texttt{MESA} are referred to as $R_\mathrm{MESA}$ hereafter.
 We also list the numerical parameters for the two simulations, including the mass of the gas in the simulation $M_\mathrm{sim}$, mass of the central point particle $M_\mathrm{IB}$, radius of the artificial core $R_\mathrm{IB}$, size of the simulation box $l_\mathrm{box}$, resolution near the stellar surface $\Delta r_\mathrm{surf}$, total number of cells $N_\mathrm{cell}$, total number of directions used in radiation transport $N_\mathrm{RT}$, and the total simulation duration $t_\mathrm{sim}$.
}
 \label{tab:sim}
 \begin{tabular}{c|ccccc|cccccccc}
\hline
\hline
 Model & & \multicolumn{3}{c}{Stellar parameters} & & \multicolumn{8}{c}{Numerical parameters}\\
 \hline
 & & $M_\mathrm{tot}$ & $L_\mathrm{bol}$ & $R_\mathrm{ross}$ & $T_\mathrm{eff}$ & $M_\mathrm{sim}$ & $M_\mathrm{IB}$ & $R_\mathrm{IB}$ & $l_\mathrm{box}$ & $\Delta r_\mathrm{surf}$ & $N_\mathrm{cell}$ & $N_\mathrm{RT}$ & $t_\mathrm{sim}$\\
 & & [$M_\odot$] & [$10^5 L_\odot$] & [$R_\odot$] & [K] & [$M_\odot$] & [$M_\odot$] & [$R_\mathrm{MESA}$] & [$R_\mathrm{MESA}$] & [$R_\odot$] & [$10^7$] & - & [yrs]\\
 \hline
 $10\, M_\odot$ & 1D \texttt{MESA} & $9.73$ & $0.94$ & $722$ & $3757$ & \multirow{2}{*}{$5.34$} & \multirow{2}{*}{$4.43$} & \multirow{2}{*}{$3\%$} & \multirow{2}{*}{$300$} & \multirow{2}{*}{$6$} & \multirow{2}{*}{$2.3$} & \multirow{2}{*}{$80$} & \multirow{2}{*}{$66$} \\
 pre-SN & 3D \texttt{AREPO} & $9.78$ & $1.09_{-0.24}^{+0.24}$ & $990_{-125}^{+132}$ & $3331_{-274}^{+376}$ & & \\
 \hline
 $20\, M_\odot$ & 1D \texttt{MESA} & $19.45$ & $2.59$ & $1132$ & $3867$ & \multirow{2}{*}{$9.52$} & \multirow{2}{*}{$10.08$} & \multirow{2}{*}{$3\%$} & \multirow{2}{*}{$300$} & \multirow{2}{*}{$8$} & \multirow{2}{*}{$2.0$} & \multirow{2}{*}{$80$} & \multirow{2}{*}{$73$} \\
 pre-SN & 3D \texttt{AREPO} & $19.60$ & $2.61_{-0.51}^{+0.58}$ & $1317_{-134}^{+111}$ & $3594_{-254}^{+220}$ & & \\
\hline
 \end{tabular}
  \end{center}
\end{table*}

\subsection{1D MESA Red Supergiant Models}

To construct the 1D initial conditions, we use the 1D stellar evolution code \texttt{MESA} \citep[version 15140; ][]{paxton2011ApJS, paxton2013ApJS, paxton2015ApJS, paxton2018ApJS, paxton2019ApJS, jermyn2023ApJS} to provide the 1D structures of RSGs.
To this end, we evolve a grid of single non-rotating massive stars at metallicity $Z=0.02$ from the zero-age main sequence (ZAMS) to the onset of core collapse (defined as the phase when the maximum infall speed inside the iron core reaches $300\, \mathrm{km\, s^{-1}}$).
We then select models with different masses and luminosities along the RSG branch.

Convection is modeled using mixing-length theory \citep[MLT;][]{bohm-vitense1958Z.Astrophys.} with a mixing-length parameter $\alpha_\mathrm{MLT}$ and the Ledoux criterion.
To account for the high convective efficiency in the RSG envelope \citep{dessart2013MNRAS, chun2018ApJ, goldberg2022ApJ}, we follow the procedure in \citet{paxton2018ApJS} and use a core mixing length parameter $\alpha_\mathrm{MLT}=1.5$ for hydrogen mass fraction $X_\mathrm{H} \leq 0.5$ and $\alpha_\mathrm{MLT}=3$ for the hydrogen-rich envelope with $X_\mathrm{H} > 0.5$.
We also include semi-convection \citep{langer1983A&A} with a semi-convection parameter $\alpha_\mathrm{SC} = 1$.
We switch on convective overshooting with step overshoot parameters of $f=0.385$ and $f_0=0.05$, as calibrated by \citet{brott2011A&A}.

For the wind mass loss, we use the \citet{vink2001A&A} recipe reduced by a factor of $3$ for effective temperature $T_\mathrm{eff} \geq 10^4$ K.
This reduction factor is chosen as suggested by both theoretical \citep{krticka2017A&A, bjorklund2021A&A, gormaz-matamala2022A&Aa} and observational studies \citep{surlan2013A&A, cohen2014MNRAS, hawcroft2021A&A}.
We use \citet{decin2024A&A} recipe for $T_\mathrm{eff} < 10^4$ K, which is comparable to \citet{beasor2020MNRAS} and likely at the lower end within the RSG wind uncertainties \citep{dejager1988A&AS, yang2023A&A, massey2023ApJ, antoniadis2024A&A}.

\subsection{Initial Conditions, 1D-3D Mapping, and Relaxation}

The central part of a giant star is extremely dense and hot, and thus requires prohibitively high spatial and temporal resolutions to simulate in 3D.
We therefore introduce an `artificial core' region, which sits at the inner $3\%\, R_\mathrm{MESA}$ of the star.
Here, $R_\mathrm{MESA}$ is the stellar radius from \texttt{MESA}.
To reconstruct the core region of the star, we use the script described in \citet{ohlmann2017A&A}.
We cut out the inner $3\%$ of the star in terms of stellar radii, place a point mass at the center, and replace the inner $3\%$ profile with a modified $\gamma=4/3$ polytrope with a proper gravitational softening length.
This is to make the core region less dense while keeping it marginally stably stratified, such that the computational power is not wasted in updating the core region.
The procedure ensures the cut-out mass is replaced by the same amount of mass and the resulting profile is in hydrostatic equilibrium.

This modified 1D profile is then mapped onto the \texttt{AREPO} 3D domain using a HEALPix grid constructed in multiple spherical shells \citep[see][]{ohlmann2017A&A}.
We place the star at the middle of the simulation box, and fill the surrounding background with low density $\rho_\mathrm{bg}=10^{-17}\, \mathrm{g\, cm^{-3}}$ and low temperature $T_\mathrm{bg}=530$ K pseudo-vacuum.
We choose a box size to be $300\, R_\mathrm{MESA}$, such that it is wide enough that any sound wave will not reach the box boundary within the simulation time.\footnote{Effectively there is no boundary, because only waves or outflows exceeding $20\, \mathrm{km\, s^{-1}}$ can reach the box boundary within $60$ years of simulation time, while the typical outgoing outflow speed is below that.}
This is done by adding hierarchical layers of boxes with lower and lower resolutions around previously constructed ones, such that almost all the mesh points are still concentrated in the star.

During the initial one dynamical timescale of the star, we continuously apply a global damping term to the momentum across the simulation box to aid relaxation following \citet{ohlmann2017A&A}.
We drop the global damping afterwards and allow the convection to set in.

\subsection{Artificial Core: Energy Source and Damping}


Within the artificial core described in the previous subsection, we apply a constant radiative luminosity that has the same value as the surface bolometric luminosity $L_\mathrm{bol,MESA}$ from \texttt{MESA}, i.e., $4\pi r^2 F_{\mathrm{rad},r}=L_\mathrm{bol,MESA}$, where $r$ is the distance from the center and $F_{\mathrm{rad},r}$ is the radial component of the radiation flux.
The intensities are constructed such that the radiation energy density is in equilibrium with the local temperature $T$ and gives the correct radial radiation flux $F_{\mathrm{rad},r}$ assuming the Eddington approximation, i.e., $I_n = caT^4/(4\pi) + 3F_{\mathrm{rad},r}(\vectoraas{n}_n\cdot\vectoraas{\hat{r}})/(4\pi)$, where $a$ is the radiation constant, $\vectoraas{n}_n$ is the unit vector along the $n^\mathrm{th}$ discretized angle for radiation transport, and $\vectoraas{\hat{r}}$ is the unit vector along the radial direction.
This acts as an inner boundary condition for the radiation transport module.
Within the artificial core, the radiation transport module is switched off and the intensities are not updated.

We further continuously apply a damping term to the momentum within the artificial core, such that $\rho\vectoraas{\dot v}=-\rho\vectoraas{v}/\tau_\mathrm{c}$, where we empirically choose a damping timescale $\tau_\mathrm{c} = 10000$ s.
We find that the damping is necessary to keep the artificial core stable throughout the simulation.
Otherwise, the core can only remain stable for about $15$ global sound-crossing timescales of the star, and then starts to deviate from hydrostatic equilibrium and generates strong sound waves that propagate outwards and dominate the dynamics.

\subsection{Hydrodynamics and Boundary Conditions}

\begin{figure}[htb!]
\centering
\includegraphics[width=\columnwidth]{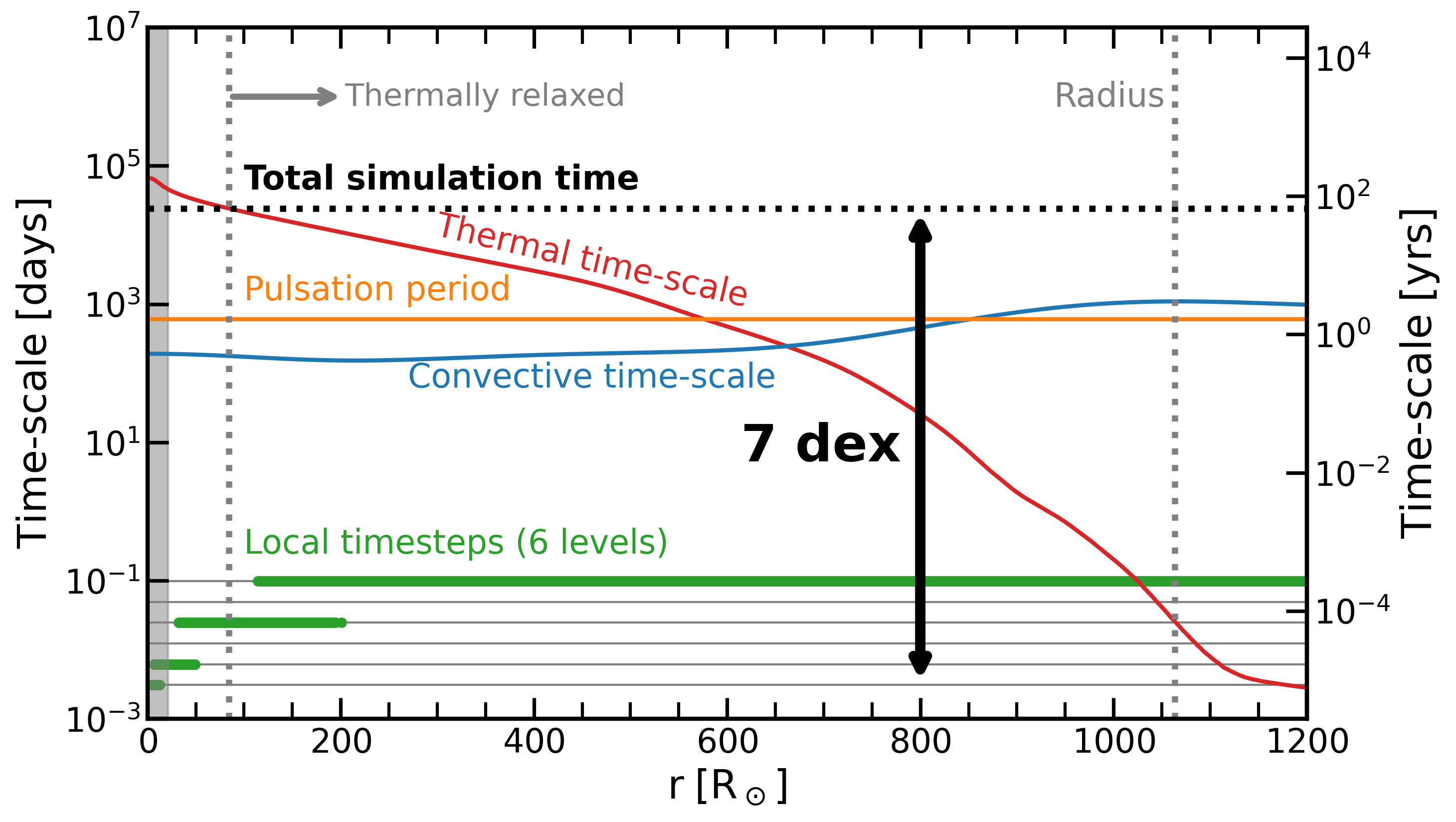}
\caption{
Our multi-scale simulation spans 7 orders of magnitude in timescales.
With the local time-stepping technique (green), we run simulations up to approximately 70 years (horizontal dotted line), capturing the convective time-scale (blue), pulsation time-scale (orange), and thermal time-scale in the upper $90\%$ of the stellar envelope.
The time-scales are calculated as in Section 5.2 in \citet{ma2025Submitt.AA}.
\label{fig:timescale}}
\end{figure}

We use the 3D moving-mesh code \texttt{AREPO} to perform radiation hydrodynamic simulations.
The hydrodynamics is solved via a second-order accurate finite-volume approach with an HLLD Riemann solver.
We solve full self-gravity using a tree-particle-mesh method.
Compared to typical 3D codes, \texttt{AREPO} performs calculations on an unstructured Voronoi moving-mesh, which allows for flexible and adaptive spatial resolutions.
\texttt{AREPO} also exploits local time-stepping, i.e. different parts of the simulation can take different timesteps grouped into a power-of-two hierarchy, which significantly reduces the computational cost for multi-scale problems.
This gives us the unique advantage of simulating the star spanning 7 orders of magnitude in time-scales, as shown in Figure~\ref{fig:timescale}.
Details for hydrodynamics, gravity solver, mesh construction, and time-stepping are presented in e.g., \citet{springel2010MNRAS}, \citet{pakmor2016MNRAS}, and \citet{weinberger2020ApJS}.

We use periodic boundary conditions for hydrodynamics and outflow boundary conditions for radiation transport on all sides to guarantee a smooth mesh construction.
The boundary conditions are not used for the gravity solver.

\subsection{Treatment of Radiation Transport}
\label{apx:method:rt}

For radiation transport, we use our recently-implemented \texttt{AREPO-IDORT} module to solve the \textit{time-independent} gray radiative transfer equations with an implicit discrete ordinates method \citep{ma2025Submitt.AA}.
This module is based on the method of \citet{jiang2021ApJS} \citep[a current radiation module in \texttt{Athena++};][]{stone2020ApJS}, but we generalize it to support local time-stepping, moving Voronoi mesh, and tabulated equation of state (EOS).
The module solves for specific intensities along discrete directions via a first-order accurate iterative finite-volume solver, updates the temperature field implicitly, and couples radiation and gas with energy and momentum exchange.
We discretize the angular space into $80$ directions, which yields a decent coverage of the $4\pi$ solid angle without introducing significant artifacts \citep[see the resolution test in figure 16 of][]{ma2025Submitt.AA}.

Since we solve the \textit{time-independent} radiation transport equations, it is more accurate to include the radiation pressure in the EOS in radiation-dominated stellar interior.
We therefore introduce a transition zone between radiation-included EOS and radiation-excluded EOS.
We include the radiation energy and radiation pressure assuming local thermal equilibrium in the EOS for density $\rho > 10^{-9}\, \mathrm{g\, cm^{-3}}$ and smoothly transition to a pure gas EOS for density $\rho < 10^{-11}\, \mathrm{g\, cm^{-3}}$ using a suppression factor multiplied onto the radiation component, such that the radiation is not included in the EOS but in the radiation transport source terms in the optically-thin regime.
In the hydrodynamic solver, the radiation force is taken into account via the EOS for $\rho > 10^{-9}\, \mathrm{g\, cm^{-3}}$ and via the radiation transport source terms for $\rho < 10^{-11}\, \mathrm{g\, cm^{-3}}$.
In the transition region between $10^{-11}$--$10^{-9}\, \mathrm{g\, cm^{-3}}$, we linearly interpolate between the radiation EOS and radiation transport source terms to calculate the radiation force.
We also include the radiation source terms in the energy equation for radiative heating/cooling.
Since the radiative diffusion timescale is orders of magnitude larger than the local timestep in the optically-thick regions and the optically-thin regions hold the largest timesteps, we only switch on the radiation transport module on the globally synchronized timesteps to save computational time.
For local timesteps that are not synchronized, we only evolve the hydrodynamics and not the radiation field in active cells.

\subsection{Resolution Criterion: Multi-shell Refinement}
\label{apx:method:refinement}

\begin{figure}[htb!]
\centering
\includegraphics[width=\columnwidth]{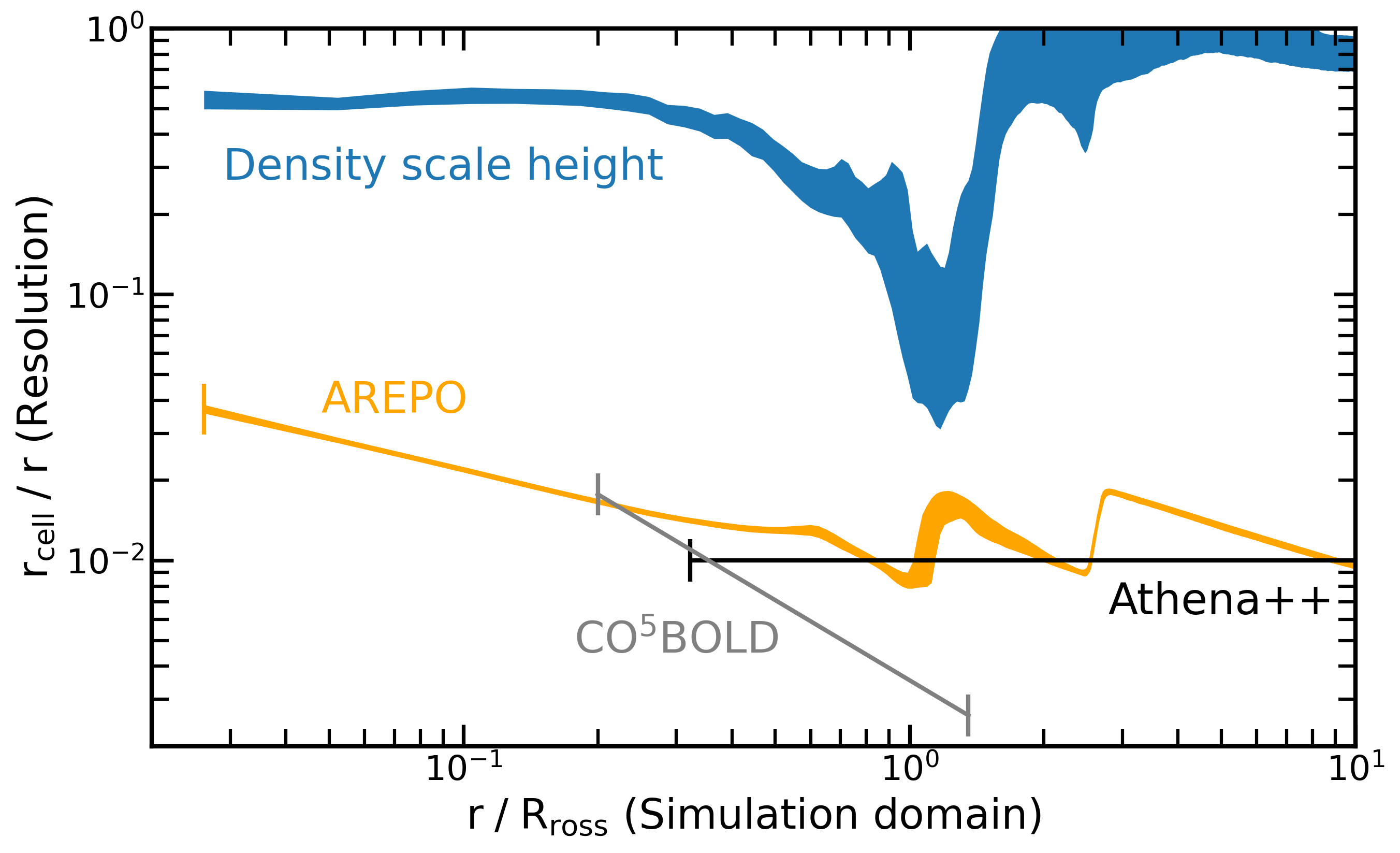}
\caption{
Spatial resolution as a function of fractional radius in the $10\, M_\odot$ simulation compared to other 3D RSG simulations.
The x axis shows the radial coordinate in the unit of stellar radius, and the y axis shows the cell size as a fraction of the local radial coordinate.
Blue indicates the $1\sigma$ range of density scale height in our $10\, M_\odot$ simulation.
An ideal resolution is to have $10$ cells per scale height.
The \texttt{CO$^5$BOLD} simulation \citep[gray line; $8\, M_\odot$ simulation in][]{freytag2024A&A} has the highest resolution, but do not simulate the inner envelope or the circumstellar environment.
The \texttt{Athena++} simulation \citep[black line; $13\, M_\odot$ simulation in][]{goldberg2022ApJ} include the circumstellar environment but not the interior.
Our \texttt{AREPO} simulation (orange line) include the deep envelope all the way to the circumstellar environment, and has a resolution comparable to \texttt{Athena++}.
This is enough to resolve the physics throughout the simulation domain except near the stellar surface.
\label{fig:res}}
\end{figure}

To resolve the stellar interior, we apply a global target mass resolution of $\Delta m = 2.7\times 10^{-7} M_\mathrm{tot}$, which guarantees approximately 10 cells per density scale height in the interior.
However, as the density drops by orders of magnitude with radial coordinate, the mass resolution criterion is insufficient to resolve the stellar surface and CSM.
We therefore apply a target cell radius of $r_\mathrm{cell} = 2\%\, R_\mathrm{MESA}$ in the spherical shell between $0.2$--$3\, R_\mathrm{MESA}$.
To resolve the stellar surface, we further apply a target cell radius of $r_\mathrm{cell,surf} = 6\, R_\odot$ ($8\, R_\odot$) when the density of the cell falls into $[10^{-7},10^{-10}]\, \mathrm{g\, cm^{-3}}$ for the $10\, M_\odot$ ($20\, M_\odot$) simulation.
To resolve the CSM structure, we then use a target cell-radius-to-distance ratio $r_\mathrm{cell}/r=0.02$ in the spherical shell between $2$--$30\, R_\mathrm{MESA}$.
To avoid numerical issues when any outflow propagates beyond that, we further apply a target mass $\Delta m = 10^{-12}\mathrm{g\; cm^{-3}}\times 4\pi\Delta r^3/3$ with $r_\mathrm{cell} = 2\%\, R_\mathrm{MESA}$ when the radial coordinate reaches beyond $2.5\, R_\mathrm{MESA}$.
The actual target cell size is taken to be the minimum value of all the criteria above, as shown in Figure~\ref{fig:res}.

\subsection{How to Determine a Steady State}
\label{apx:method:steadystate}

\begin{figure*}[htb!]
\centering
\includegraphics[width=\textwidth]{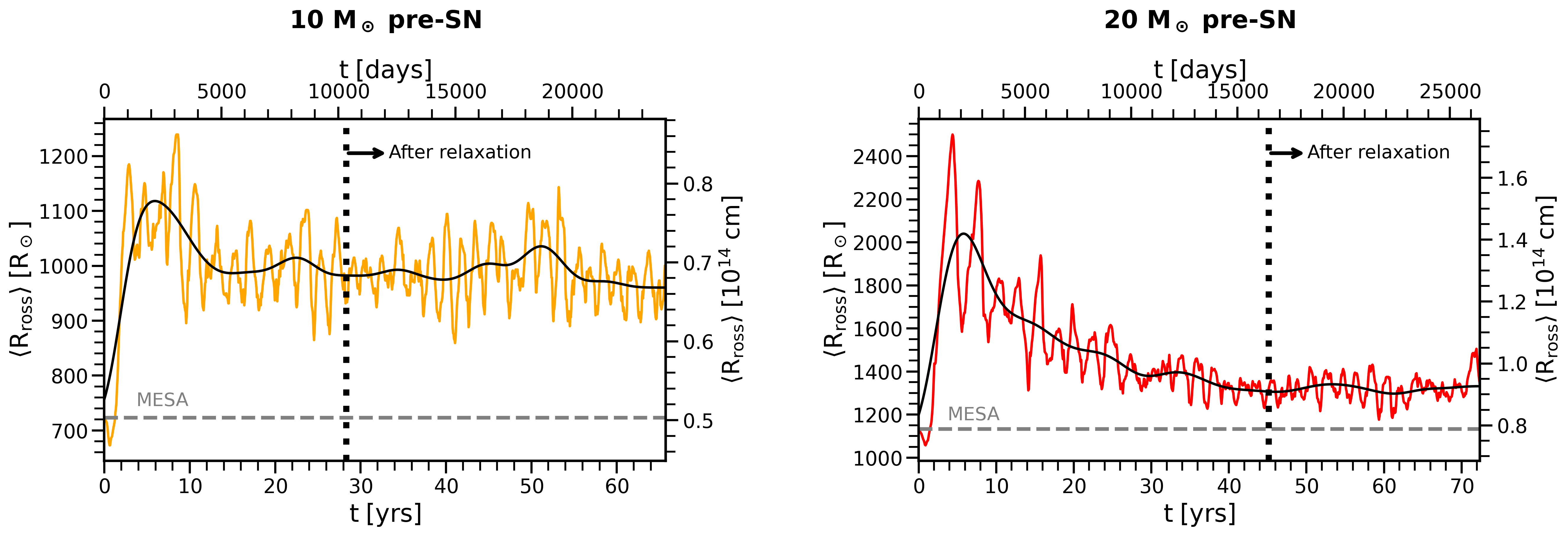}
\caption{
Spherically-averaged Rosseland radii vary around constant values after the relaxation phase.
\label{fig:0d:steadyradius}}
\end{figure*}

\begin{figure*}[htb!]
\centering
\includegraphics[width=\textwidth]{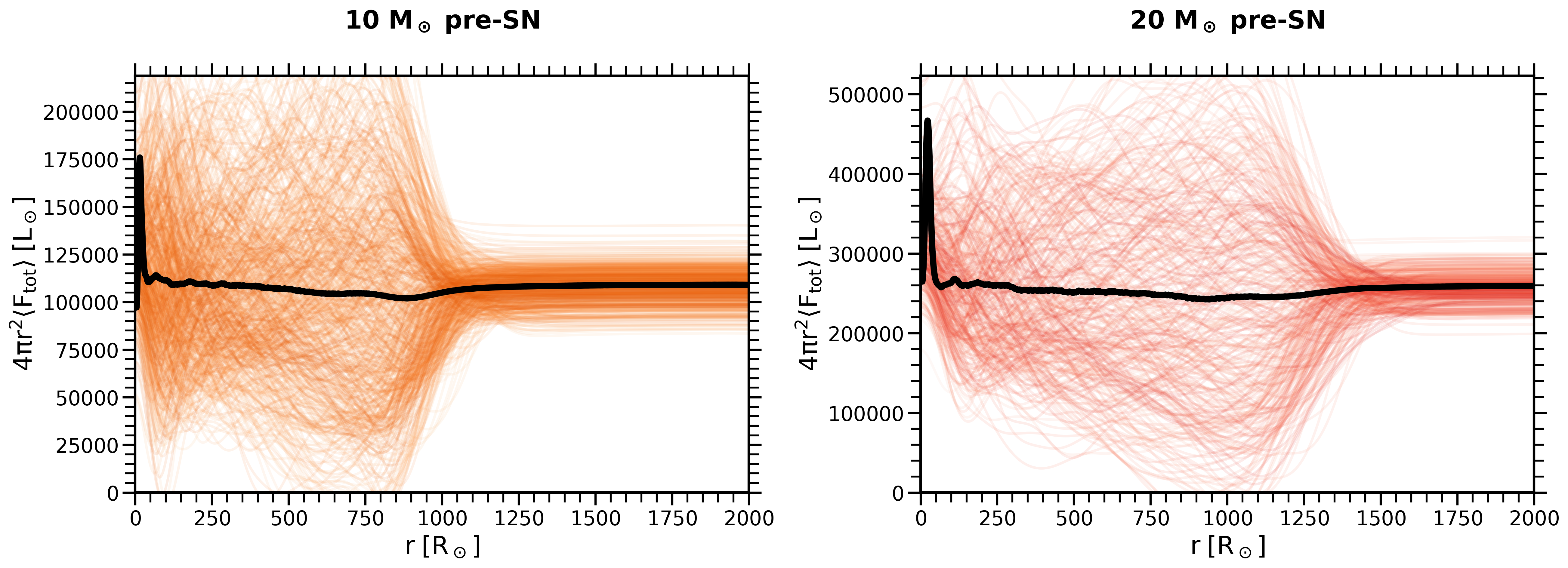}
\caption{
Time-averaged total energy flux is spatially constant above the artificial core during the steady states for both simulations.
The colored lines show the total energy flux at different times, and the black line shows the time-averaged values of the colored lines.
\label{fig:0d:steady}}
\end{figure*}

We determine the steady state based on three general considerations: (1) The mass ejection is not influenced by the initial transient ejection anymore; (2) The global quantities (e.g., luminosity and radius) do not show an increasing/decreasing trend \citep{ahmad2023A&A}; (3) The time-averaged total radial energy flux does not vary with radius \citep{goldberg2022ApJ, goldberg2025arXive-prints}.
We find the first criterion is the most stringent one, and we also check the other two are satisfied during our defined steady state.

The first criterion is illustrated in the third row of Figure~\ref{fig:0d:tvar}, where we use the spherically-averaged density contour at $10^{-16}\; \mathrm{g\; cm^{-3}}$ as an indicator.
This density contour line shows a general decreasing trend after the first initial transient ejections, until the mass ejections are not influenced by the fallback material anymore and grow again.
We determine the onset of the steady state as the time corresponding to the minimum point of this density contour line.

We also qualitatively check the second and third criteria.
The bolometric luminosity and spherically-averaged Rosseland radius vary around constant values, as shown in Figure~\ref{fig:0d:tvar} and Figure~\ref{fig:0d:steadyradius}.
The time-averaged total energy flux is also constant as a function of radial coordinate down to the inner boundary (Figure~\ref{fig:0d:steady}), indicating an energy equilibrium state.

\subsection{Method for Analyzing the 3D Data}
\label{apx:method:analysis}

For all the spherically-averaged quantities shown in this work, we use a radial binning method to analyze the \texttt{AREPO} simulation data.
We average the quantities of all the cells falling inside thin spherical shells of radius $r$ and thickness $\delta R = R_\mathrm{MESA}/1000$.
In this work, the spherically-averaged density and energy fluxes are weighted by volume, whereas all other quantities are weighted by mass.
The spherically-averaged quantity $q$ weighted by volume is $\langle q\rangle_V (r) \equiv \left[\sum_{|r_i-r|<\delta R/2}q_i\Delta V_i\right]/\left[\sum_{|r_i-r|<\delta R/2}\Delta V_i\right]$, where $\Delta V_i$ is the volume enclosed in each cell, and the sum is performed over all the cells $i$ whose radial distance $r_i$ from the center falls into the $(r-\delta R/2,r+\delta R/2)$.
The spherically-averaged quantity weighted by mass is $\langle q\rangle_m (r) \equiv \left[\sum_{|r_i-r|<\delta R/2}q_i\Delta m_i\right]/\left[\sum_{|r_i-r|<\delta R/2}\Delta m_i\right]$, where $\Delta m_i$ is the mass enclosed in cell $i$.

To obtain the fundamental parameters of the star from our simulations, we adopt a more specific approach (as in \citealt{ma2025Submitt.AA}).
The bolometric luminosity $L_\mathrm{bol}$ is calculated by taking the radial component of the radiation flux $F_{\mathrm{rad},r}$ from the simulation output and performing spherical average for $4\pi r^2 F_{\mathrm{rad},r}$ within the spherical shell in $4$--$5\, R_\mathrm{MESA}$\footnote{It is more intuitive to calculate the bolometric luminosity by integrating the normal radiation flux over box boundaries. However, since the box boundary is far away from the star with very low resolution, it is more accurate to calculate the luminosity closer to the star.}.
The spherically-averaged Rosseland optical depth is obtained by summing up the optical depth from the outer boundary to the distance $r$, i.e. $\langle \tau_\mathrm{ross} \rangle (r)\equiv \sum_{r_i>r}\left[\kappa_\mathrm{R}\Delta m/(4\pi r^2)\right]_i$.
Here, for each cell $i$, $\kappa_\mathrm{R}$ is the Rosseland opacity, $\Delta m$ is the mass enclosed in the cell, and $r$ is the distance of the cell from the stellar center.
We define the Rosseland radius $\langle R_\mathrm{ross}\rangle$ as the radius where $\langle \tau_\mathrm{ross}\rangle (r=\langle R_\mathrm{ross}\rangle) = 1$.
The spherically-averaged effective temperature $\langle T_\mathrm{eff}\rangle$ is defined via the Stefan-Boltzmann law, i.e. by finding the temperature that satisfies $L_\mathrm{bol}=4\pi \langle R_\mathrm{ross}\rangle^2\sigma \langle T_\mathrm{eff}\rangle^4$, where $\sigma$ is the Stefan-Boltzmann constant.



\section{Identifying the Dominant Pulsation Mode}
\label{apx:analyses:fourier}

\begin{figure*}[htb!]
\centering
\includegraphics[width=\textwidth]{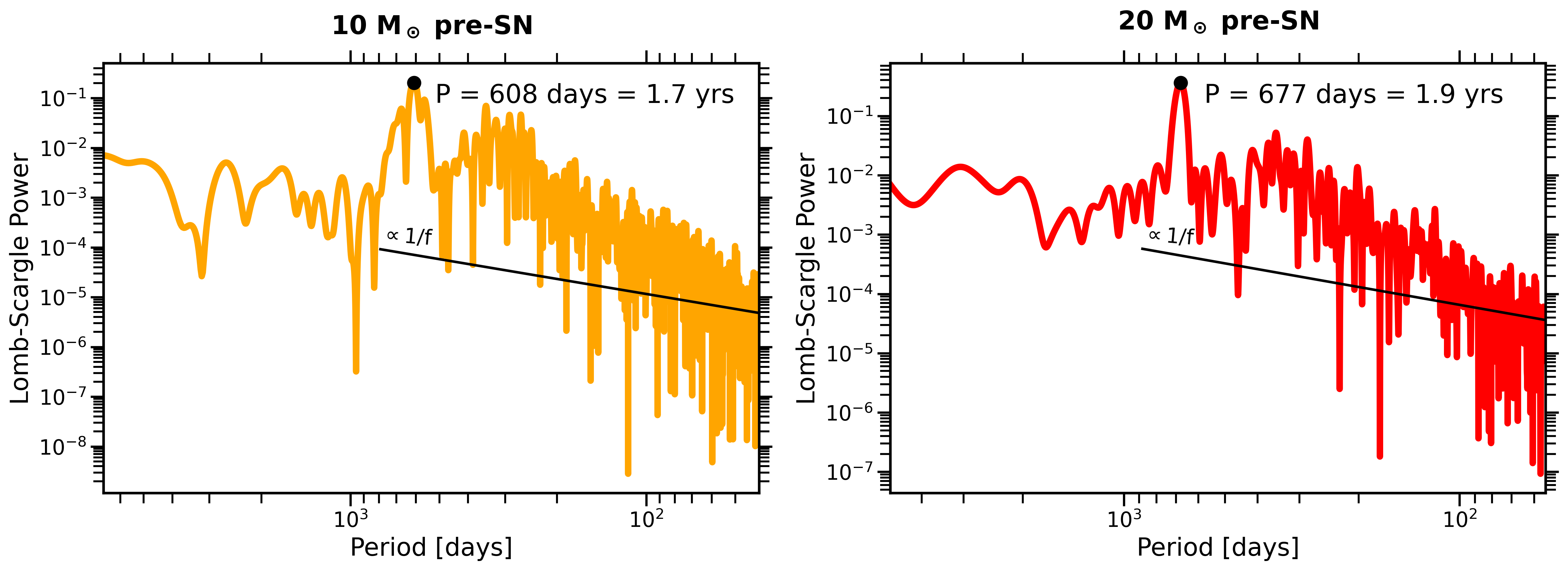}
\caption{
Power spectrum of the simulated lightcurve in Figure~\ref{fig:0d:tvar}.
We obtain the period of the dominant pulsation mode in our simulations by identifying the maximum peak in the power spectrum.
The high-frequency part of the power spectrum falls off more steeply than the $1/f$ trend observed in normal RSGs \citep[where $f$ is the frequency;][]{kiss2006MNRAS}, but closer to simulated yellow supergiants \citep{goldberg2025arXive-prints} and other observed massive stars \citep{bowman2023Ap&SS}.
\label{fig:0d:fourier}}
\end{figure*}

\begin{figure*}[htb!]
\centering
\includegraphics[width=0.8\textwidth]{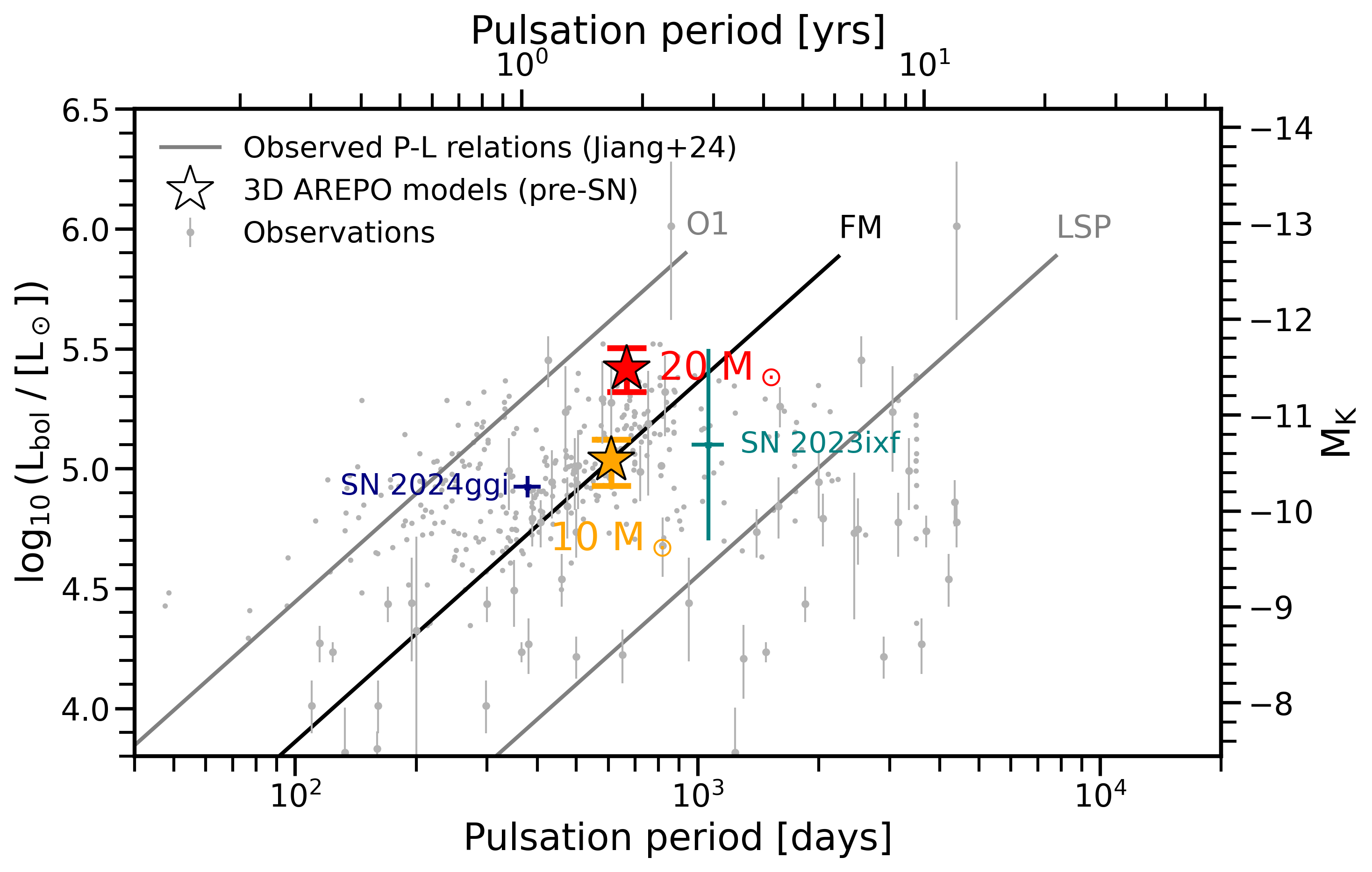}
\caption{
The dominant pulsation periods in our 3D simulations agree with the fundamental modes of RSGs according to the empirical period-luminosity relation.
The pulsation periods of the 3D simulations are found by identifying the dominant peak in the power spectra (Figure~\ref{fig:0d:fourier}) of the simulated lightcurves (Figure~\ref{fig:0d:tvar} top row).
We plot the observed period-luminosity relations of RSGs from \citet{jiang2024IAUSymp.} in solid lines, separated into the fundamental mode (FM), first overtone (O1), and long secondary period (LSP).
In gray scatter dots, we show the observed RSG population in the Galaxy \citep{chatys2019MNRAS}, M31 \citep{soraisam2018ApJ}, and M33 \citep{ren2019ApJS}.
All the observed data points and observed relations are in absolute K-band magnitude $M_K$, and we convert them into bolometric luminosities $L_\mathrm{bol}$ following the reddening relation in \citet{massey2009ApJ} assuming an overall effective temperature of $3800$ K, which is a crude approximation.
We highlight the pulsation periods identified from the pre-explosion lightcurves of two interacting SN progenitors, SN 2023ixf \citep{kilpatrick2023ApJ, soraisam2023ApJ, qin2024MNRAS, xiang2024Sci.ChinaPhys.Mech.Astron.} and SN 2024ggi \citep{xiang2024ApJ}.
\label{fig:0d:plr}}
\end{figure*}

By taking the standard Lomb-Scargle periodogram \citep{lomb1976Ap&SS, scargle1982ApJ} of the simulated lightcurves in the top row of Figure~\ref{fig:0d:tvar}, we obtain the power spectra of the lightcurves (Figure~\ref{fig:0d:fourier}) and identify the dominant period with the maximum power.
In Figure~\ref{fig:0d:plr}, we compare the dominant periods of both simulations with the period-luminosity relations of observed RSG populations from \citet{jiang2024IAUSymp.}.
We find that the dominant pulsation modes in our simulations fit reasonably well with the fundamental modes according to the period-luminosity relations.
This agrees with most theoretical and observational results that suggest Galactic RSGs pulsate predominantly in the fundamental mode \citep{heger1997A&A, kiss2006MNRAS, yang2012ApJ, ren2019ApJS, joyce2020ApJ, jiang2024IAUSymp., suzuki2025arXive-prints}, however see \citet{guo2002ApJ}.

In Figure~\ref{fig:0d:plr}, we also mark the pulsation period and luminosity inferred from pre-explosion images of two interacting SN progenitors, SN 2023ixf \citep{kilpatrick2023ApJ, soraisam2023ApJ, qin2024MNRAS, xiang2024Sci.ChinaPhys.Mech.Astron.} and SN 2024ggi \citep{xiang2024ApJ}.
By comparing the pulsation period of SN 2023ixf progenitor with our simulations and the empirical period-luminosity relations, we support the claim of \citet{soraisam2023ApJ} and \citet{hsu2024arXive-prints} that the pulsation period of SN 2023ixf progenitor fits better with a luminous RSG of $\sim 20\, M_\odot$.
The apparent pulsation period of the SN 2024ggi progenitor instead favors a very low-mass RSG, but whether this inferred pulsation period is true is debated \citep{laplace2025arXive-prints}.

\section{Additional Discussion}
\label{apx:disc}

\subsection{Comparison with Other 1D Models}


Current analyses of observed Type II SN lightcurves and spectra rely heavily on comparison with 1D SN modeling, which is sensitive to the 1D progenitor model and the CSM profile \citep[e.g.,][]{dessart2013MNRAS, dessart2017A&A, morozova2017ApJ, dessart2019A&A, goldberg2019ApJ, goldberg2020ApJa, moriya2018MNRAS, boian2020MNRAS, moriya2023PASJ}.
We find that our 3D simulations differ from the 1D models in many aspects.
The differences in 1D-averaged quantities include different CSM density profile, the presence of large-amplitude pulsations, and larger radii.
We discuss those differences separately in this subsection.

\subsubsection{CSM Density Profile}

When interpreting the observed Type II SN lightcurves and spectra, most works assume that the CSM follows a steady-wind density profile governed by $\rho\propto r^{-2}$ \citep[e.g.,][]{morozova2017ApJ, boian2020MNRAS, jacobson-galan2024ApJa}.
However, it has been shown that the inferred CSM density, mass and radial extent are sensitive to the density structure assumed, where variations include wind acceleration \citep{moriya2017MNRAS, moriya2018MNRAS, dessart2025A&A} and extended atmosphere \citep{dessart2017A&A, soker2021ApJ, dessart2023A&A, fuller2024OJAp}.

Our 3D simulations broadly support the idea of an extended atmosphere, but the detailed physics and the CSM density structure are different from previous works.
\citet{dessart2017A&A} and \citet{dessart2023A&A} used an ad hoc atmospheric extension by assuming an exponentially-decaying CSM density profile as a function of the distance from the stellar surface.
\citet{soker2021ApJ} proposed an `effervescent zone' where bound clumps are ejected by stellar activity and fall back.
The radial extent is determined by the balance of gravity, wind drag and radiation force, and the density profile can be uncertain \citep{soker2023RAA}.
\citet{fuller2024OJAp} proposed a `chromosphere' model where shock waves launched by transonic convection can support a dense atmosphere that extends to the dust-formation radius and radiation acts on dust to drive a wind.

Our simulations support part of the \citet{soker2021ApJ} model and part of the \citet{fuller2024OJAp} model:
We find large-scale clumps episodically lifted by pulsations and shaped by convection in agreement with \citet{soker2021ApJ}, but the wind is not present in the immediate surrounding of the star.
The wind is launched later on when some clumps reach far enough to cool down and form dust, in agreement with \citet{fuller2024OJAp}.
The density profile is steeper than proposed by \citet{fuller2024OJAp}, because we find that pulsation instead of convection is the main driver of the shock waves, although convection can break the shock waves into multiple curved shock fronts.
The spherically-averaged shock wave speeds are nearly constant due to semi-regular pulsation, instead of a Gaussian distribution due to convection adopted in \citet{fuller2024OJAp}.
Generally, our simulations suggest a `two-zone model' composed of a periodic-shock-supported atmosphere attached to a dust-driven wind as described in Section~\ref{sec:result:agb}, which is effectively a combination of the \citet{soker2021ApJ} model and the \citet{fuller2024OJAp} model.

\subsubsection{Large-amplitude Pulsation}

RSGs are known to be large-amplitude semi-regular radial pulsators from both observations \citep[e.g][]{kiss2006MNRAS, soraisam2018ApJ, jiang2024IAUSymp.} and theories \citep[e.g.,][]{guo2002ApJ, joyce2020ApJ, bronner2025arXive-prints, sengupta2025arXive-prints, suzuki2025arXive-prints}.
However, large-amplitude radial pulsations are non-linear, so their amplitudes are difficult to predict.
Especially for pre-explosion RSGs, 1D models suggest that their high $L/M$ ratio amplifies the pulsation amplitude \citep{heger1997A&A, joyce2020ApJ, bronner2025arXive-prints, suzuki2025arXive-prints} and may even trigger a `superwind' or mass loss \citep{yoon2010ApJa, clayton2018, sengupta2025arXive-prints}.
However, all the current 1D models suffer from significant numerical damping and the surface cooling is not correctly captured \citep{heger1997A&A, clayton2018, joyce2020ApJ, bronner2025arXive-prints, suzuki2025arXive-prints}, which are important for predicting the growth and damping rates of pulsation.

In this work, we self-consistently predict the pulsation amplitudes (Figure~\ref{fig:0d:tvar}).
We also find the dominant pulsation periods of our simulations agree with the fundamental mode expected for RSGs (Figure~\ref{fig:0d:plr}).
If true, this means the fundamental mode dominates over other higher-order modes in pre-explosion RSGs.
This leads to different density structures especially near the stellar surface and different radii when the star explodes, which is expected to create some diversity in SN lightcurves \citep{goldberg2020ApJ, bronner2025arXive-prints}.
We also find that, contrary to the suggestions in \citet{heger1997A&A}, \citet{yoon2010ApJa}, \citet{clayton2018} and \citet{sengupta2025arXive-prints}, the pulsations are not strong enough to unbind any material without the help of molecules, dust, or magnetic fields.

\subsubsection{Larger Radii and Convective Efficiency}
\label{apx:disc:radius}

\begin{figure*}[htb!]
\centering
\includegraphics[width=0.8\textwidth]{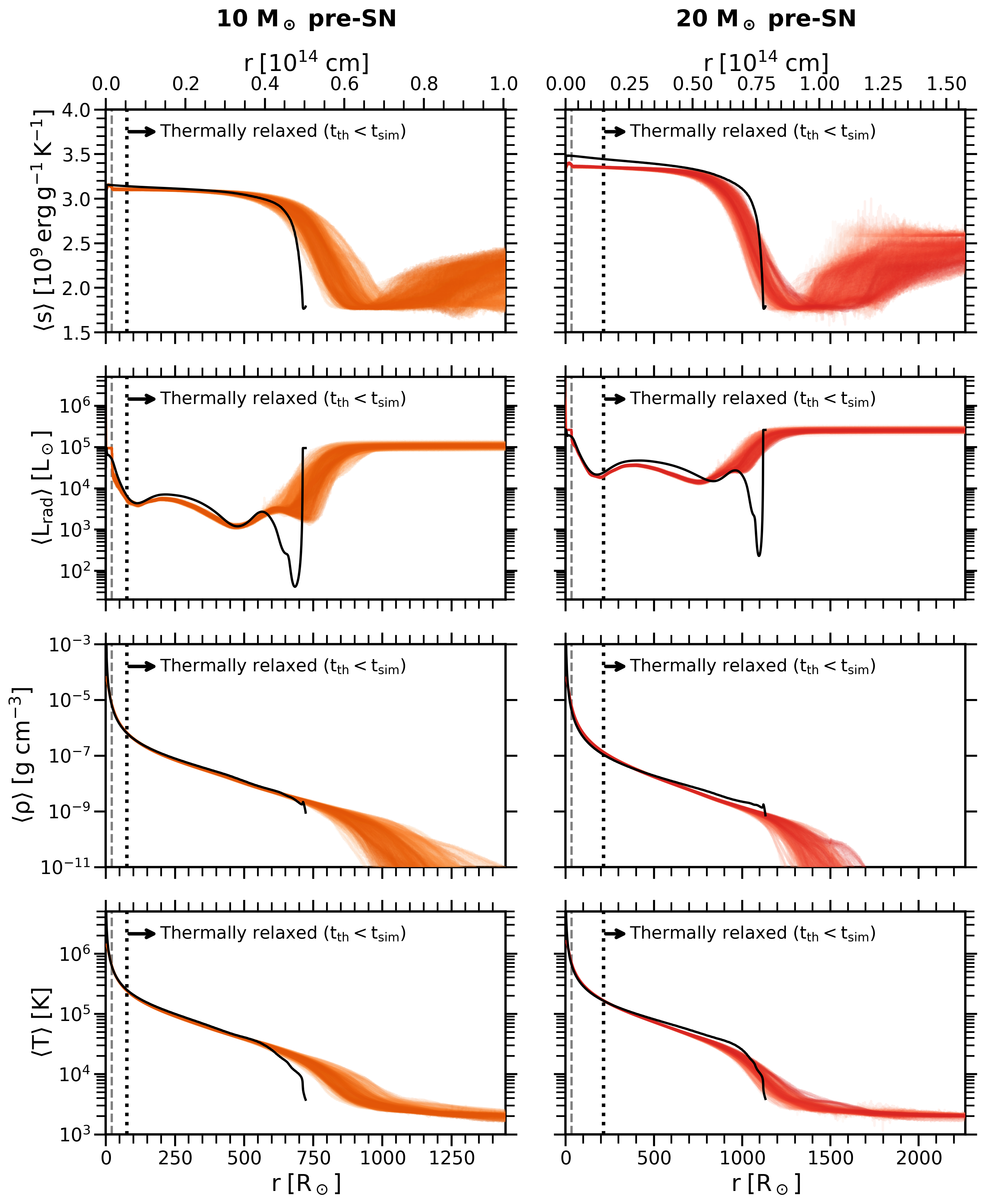}
\caption{
Comparison of 1D \texttt{MESA} profiles (black) and spherically-averaged 3D \texttt{AREPO} profiles (colored).
Different colored curves show the spherically-averaged 3D profiles at different times.
From top to bottom, we show the spherically-averaged specific entropy $\langle s \rangle$, radiative luminosity $\langle L_\mathrm{rad}\rangle$, density $\langle\rho\rangle$, and temperature $\langle T\rangle$.
The profiles on the right of the vertical dotted lines are thermally relaxed, i.e. the thermal timescale is smaller than the simulation duration.
In our simulations, the deep entropy profile is nearly flat (first row), and radiation only transports a small fraction of energy (second row), which agrees with the expectation of efficient convection.
As pointed out in \citet{goldberg2022ApJ} and \citet{chiavassa2024LRCA}, this was correctly simulated in \texttt{Athena++} \citep{goldberg2022ApJ} but not in \texttt{CO$^5$BOLD} \citep{chiavassa2011A&Aa}.
As shown in all panels, the deep interior of the 3D simulations agree very well with the 1D profiles.
This direct 1D-3D agreement inside the star was not achieved in previous simulations \citep{chiavassa2011A&Aa, goldberg2022ApJ}.
\label{fig:1d:1d3d}}
\end{figure*}

The RSG radii are extremely sensitive to the convective efficiency in the envelope: 
To transport the same amount of energy, if convection is more efficient, then the radiation flux is lower and the temperature gradient is flatter, which yields larger surface temperature and therefore smaller radius.
In 1D stellar models, convection is commonly described using mixing-length theory \citep[MLT;][]{bohm-vitense1958Z.Astrophys.}.
The convective efficiency is controlled by $\alpha_\mathrm{MLT}$, the ratio between mixing length and local pressure scale height.
Larger $\alpha_\mathrm{MLT}$ means more efficient convective energy transport, resulting in more compact RSGs \citep[e.g.,][]{henyey1965ApJ, stothers1995ApJ, goldberg2022ApJ}.

We find that the averaged radii in our 3D simulations are slightly larger than the 1D models with $\alpha_\mathrm{MLT}=3$ (by $40\%$ for $10\, M_\odot$ simulation and $20\%$ for $20\, M_\odot$ simulation).
To better constrain the convective efficiency, we compare the averaged entropy profiles from our 3D simulations to 1D initial conditions in Figure~\ref{fig:1d:1d3d}.
The 3D entropy profiles are flatter than 1D \texttt{MESA} profiles in the interior of the envelope, but the entropy drop in the superadiabatic layer is less steep than 1D \texttt{MESA} profiles.
This indicates that convection in our 3D simulations is more efficient than 1D \texttt{MESA} models in the interior of the envelope, but is less efficient than 1D \texttt{MESA} models in the superadiabatic layer.
We therefore suggest that the mixing-length parameters may be better described by $\alpha_\mathrm{MLT} \gtrsim 4$ in the efficient convection zone below the hydrogen opacity bump.
However, near the surface in the superadiabatic layer, we either favor $\alpha_\mathrm{MLT} \lesssim 2$ or the turbulent pressure needs to be taken into account to inflate the envelope.
In our simulations, the inflated superadiabatic layer is the main reason for larger radii compared to 1D models with $\alpha_\mathrm{MLT}=3$.
Our findings for the convective efficiency are fully consistent with the results of 3D RSG simulations with \texttt{Athena++} \citep{goldberg2022ApJ}.

However, the large radii found in our simulations potentially contradict with the results of \citet{dessart2013MNRAS}, where they argued for compact pre-explosion RSGs with $\alpha_\mathrm{MLT}=3$ because SN radiation from RSGs with large radii will have weaker cooling and remain blue for too long.
This controversy may be complicated in two ways.
First, it is not clear whether the Rosseland radius determined in the simulation is the same as the stellar radius the SN shock is sensitive to \citep{dessart2013MNRAS}.
Second, clumping may have a counter effect on the color evolution \citep{dessart2018A&A}, thereby softening the degeneracy.
A detailed analysis of the temperature gradient and turbulent pressure in our 3D simulations is needed to assess the actual convective efficiency, as in e.g., \citet{goldberg2022ApJ}.

\subsection{Comparison with Other 3D Radiation Hydrodynamic Simulations}

\begin{figure*}[htb!]
\centering
\includegraphics[width=\textwidth]{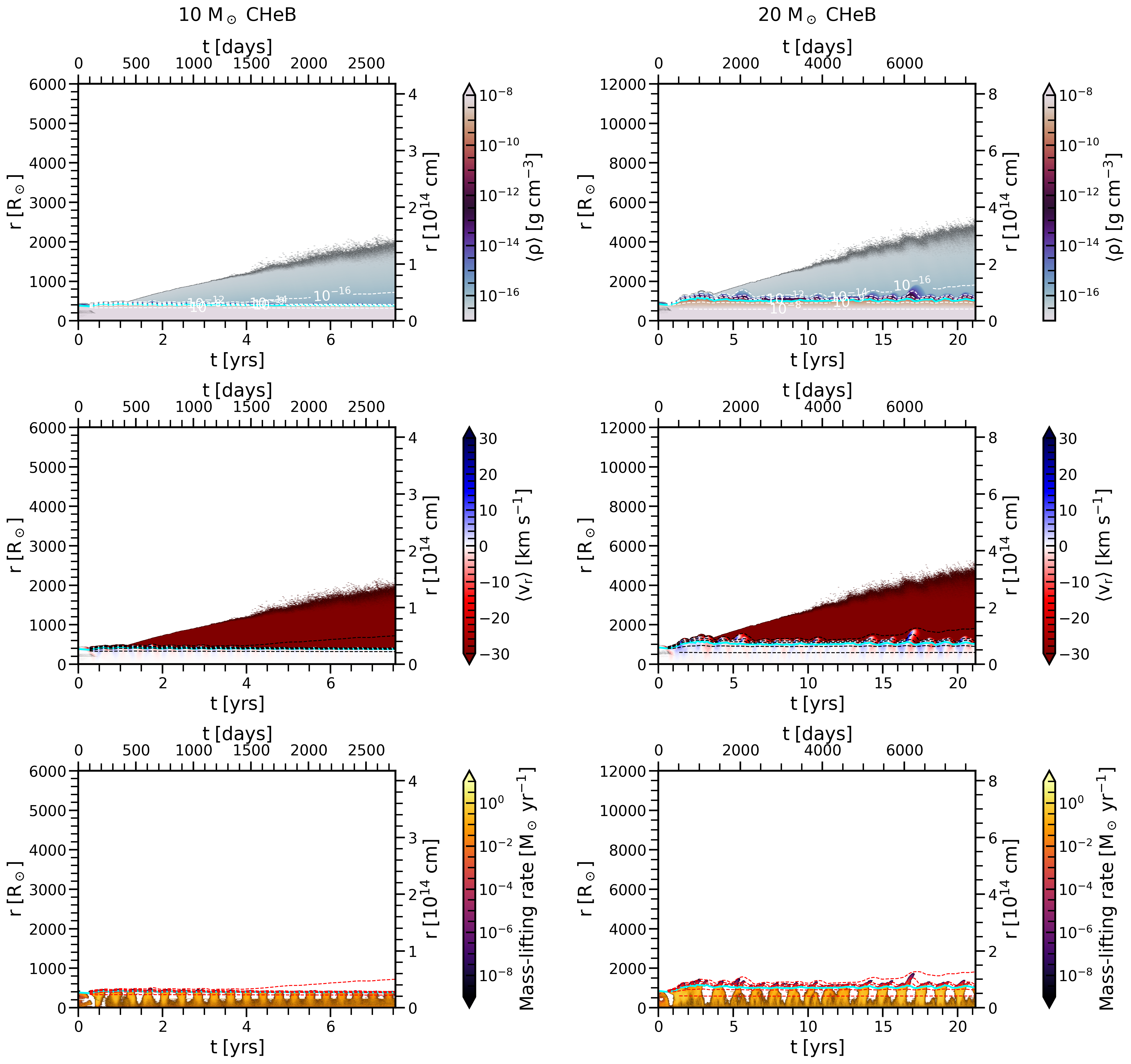}
\caption{
Core helium burning RSGs have very weak pulses and do not produce the CSM in our 3D simulations.
Similar to Figure~\ref{fig:0d:tvar}, but for two additional 3D simulations with lower luminosities than the simulations we present elsewhere in this work, taken during core helium burning.
The mass-lifting rate is defined as $4\pi r^2\langle\rho\rangle\langle v_r\rangle$.
\label{fig:cheb:1d:tr}}
\end{figure*}


RSGs have been simulated with 3D radiation hydrodynamics by e.g., \citet{freytag2002AstronomischeNachrichten, freytag2024A&A} and \citet{chiavassa2011A&Aa} using the \texttt{CO$^5$BOLD} code \citep{freytag2012JCoPh} and by \citet{goldberg2022ApJ} using the \texttt{Athena++} code \citep{stone2020ApJS}.
For a review summarizing the simulation results from both codes, see \citet{chiavassa2024LRCA}.
Other multi-dimensional hydrodynamic simulations also exist for RSGs especially in the context of transients \citep[e.g.,][]{leung2020ApJ, antoni2022MNRAS}, albeit without detailed radiation transport.

In this comparison, one major puzzle is why propagating shocks waves and extended CSM are clearly present in \texttt{CO$^5$BOLD} simulations \citep{kravchenko2019A&A, freytag2024A&A} and our \texttt{AREPO} simulations (this work), but are rather weak in \texttt{Athena++} simulations \citep{goldberg2022ApJ}.
\citet{fuller2024OJAp} suggested that this discrepancy may be due to a numerical transient in \texttt{Athena++} simulations, which ejects material during the first relaxing phase that later on falls back and quenches the shocks propagating outwards.
This numerical transient is also present in our simulations (Figure~\ref{fig:0d:tvar}) and especially strong in our $20\, M_\odot$ simulation which quenches the material ejection up to $45$ years.
\citet{fuller2024OJAp} also suggested that the recombination energy is neglected in \texttt{Athena++} simulations, which may result in weaker pulses.
Indeed, the recombination energy is included in our \texttt{AREPO} simulations and in \texttt{CO$^5$BOLD}, but not in \texttt{Athena++} where they used an ideal gas.

However, we point out here that this discrepancy in producing shocks and CSM may be due to physical reason related to the growth of strong pulsations.
We have performed two additional simulations at lower luminosities, using the 1D \texttt{MESA} models at core helium burning stage instead of pre-explosion stage.
We find that they also have weak pulsations and do not produce any extended CSM, as shown in Figure~\ref{fig:cheb:1d:tr}.
This behavior is very similar to the \texttt{Athena++} simulations and indicates a physical origin.
One possible explanation is that the $L/M$ ratios used in the \texttt{Athena++} simulations and our two additional core-helium-burning simulations are too low to foster growth of the pulsation amplitudes.
Indeed, 1D models suggest that strong pulsations are only present for RSGs with large $L/M$ ratios \citep{heger1997A&A, yoon2010ApJa, clayton2018, joyce2020ApJ, laplace2025arXive-prints, sengupta2025arXive-prints, suzuki2025arXive-prints}.

Besides differences in physical parameters, we have also made improvements both by expanding the simulation domain and including more physics compared to previous 3D simulations.
\texttt{Athena++} simulations only include a wedge of the star, but can extend the simulation domain to more than 10 times the stellar radius to capture the CSM structure, outflow, and shock breakout \citep{goldberg2022ApJ, goldberg2022ApJa}.
\texttt{CO$^5$BOLD} RSG simulations simulate the entire $4\pi$ sphere to capture the global convective structure and facilitate synthetic observations \citep[e.g.,][]{chiavassa2009A&A, ma2024ApJL}, but are limited to small box sizes about 3 times larger than the star \citep[but see][that extends the box for AGB stars]{freytag2023A&A}.
In our \texttt{AREPO} simulations, we achieve both points and therefore have the capability to simulate the global $4\pi$ convection and the CSM and outflow in a single simulation, which made this work possible.
Furthermore, the local time-stepping technique in \texttt{AREPO} gives us the unique advantage of including the deep envelope in our simulations to make sure the deep physical conditions match the initial 1D profiles from stellar evolution code (Figure~\ref{fig:1d:1d3d}), which was not achieved in previous simulations.
For reference, our effective inner boundary is at $3\%$ stellar radius in \texttt{AREPO}, while the inner boundary is at $20\%$ stellar radius in \texttt{CO$^5$BOLD} \citep{chiavassa2011A&Aa, ahmad2023A&A} and $30\%$--$50\%$ in \texttt{Athena++} \citep{goldberg2022ApJ}.
We have performed test simulations showing that moving the boundary of the artificial core too high up in the envelope ($10\%$ stellar radius) will result in weaker pulsations, which is especially important for this study.
For the included physics, \texttt{CO$^5$BOLD} simulations did not include radiation pressure or self-gravity, which are important for massive stars and loosely-bound envelopes.
\texttt{Athena++} simulations included radiation pressure and self-gravity, but used an ideal gas for the equation of state, thereby neglecting recombination energy, which is important for ejecting the envelope material.
We have included all this physics in our \texttt{AREPO} simulations.
Our 3D RSG simulations are also the first to experiment with the effects of dust, albeit under crude approximations.

There are also certain aspects where previous simulations perform better than ours.
Our simulations and \texttt{Athena++} RSG simulations are so far limited to gray radiation transport, but it has been shown in \texttt{CO$^5$BOLD} simulations that non-gray effects create a steeper temperature gradient near the surface and in the atmosphere, which is important for the stellar spectrum and measuring stellar radii \citep{chiavassa2011A&Aa}.
Furthermore, resolving the stellar photosphere is important to obtain the correct luminosity and cooling rate.
Our cell size near the stellar surface is about $0.8\%$ of the stellar radii, which is comparable to \texttt{Athena++} simulations ($1\%$ stellar radii) but not as resolved as the latest \texttt{CO$^5$BOLD} simulations ($0.4\%$--$0.5\%$ stellar radii), as shown in Figure~\ref{fig:res}.
In the optically-thick regime, we separate the radiative heating/cooling provided by radiation transport from the radiation pressure and radiation energy provided by the equation of state, which presumably yields higher-order accuracy in force balance but is not self-consistent.
The \texttt{Athena++} simulations calculate all the radiation quantities and coupling terms from the time-dependent radiation transport, which is physically more self-consistent.
In addition, due to the transition from a radiation-included equation of state to a radiation-excluded one, we introduce an artificial temperature jump in our simulation at $10^{-11}\, \mathrm{g\, cm^{-3}}$ in the atmosphere, which is not present in other simulations.
Future efforts will be devoted to solving these minor issues.

\subsection{Simulation Caveats}
\label{apx:discussion:caveat}

In this subsection, we summarize the caveats known to us for future improvements.

In our simulations, the first strong pulsations are triggered by mapping from 1D to 3D, which is a numerical artifact.
This numerical transient launches a strong ejection that affects subsequent CSM structure near the star for $30$--$45$ years until most of the material has fallen back (Figure~\ref{fig:0d:tvar}).
The same issue was also found in \texttt{Athena++} simulations \citep{goldberg2022ApJ}.
The way we deal with this is to run the simulations long enough such that the effects of the initial numerical transient die down.
However, the dust-driven outflow launched by the numerical transient still continues to propagate outwards and affect the long-range atmospheric structure, which is so far not properly treated.

This also raises the question of whether the strong pulsations seen in our simulations are real or due to numerical artifacts.
To address this question, we run two additional simulations at lower luminosities by taking 1D \texttt{MESA} profiles not at pre-explosion but at core helium burning stage.
As shown in Figure~\ref{fig:cheb:1d:tr} in comparison with Figure~\ref{fig:0d:tvar}, the 3D core-helium-burning models only have weak pulsations and do not create an extended CSM.
This means the strong pulsations in the pre-explosion models are likely not due to numerical artifacts but are related to enhanced luminosity and different stellar structures.
In addition, without a physical mechanism to continuously inject energy into pulsations, the radial pulsations are expected to be damped, e.g.,\ by convection at the timescale of several years \citep{macleod2023ApJ}.
Instead in our $10\, M_\odot$ simulation, the pulsation grows stronger again after $30$ years, indicating self-excited pulsations.
However, the exact pulsation amplitude can be influenced by numerical resolution and uncertain non-gray effects.

Even though we have included a larger portion of the convective envelope than all previous simulations, we still cannot reach the bottom of the convective envelope.
Our artificial core (or the effective inner boundary) still extends in the convective zone.
Given the non-locality of convection in these RSG envelopes, the artificial core will likely affect the convection higher in the envelope.
The effects can only be quantified by performing test simulations and putting the boundary of the artificial core at a different radius, but those simulations are currently too expensive to run until they reach a steady state.
We have performed short test simulations that suggest a higher inner boundary, at $10\%$ of the stellar radius, yields weaker pulses and weaker convection.
Since we damp the velocities in the artificial core, we also do not fully conserve the total energy, momentum, or angular momentum, but we conserve the total mass to near machine precision.
Furthermore, even if we can simulate the entire convective envelope, the deep thermal timescale is still one order of magnitude larger than the plausible simulation time, so the deep envelope will not be fully thermally relaxed.
It is reasonable to assume that, by starting from a 1D model from a stellar evolution code, the deep envelope is not far away from the actual thermally-relaxed profile, but this is difficult to test.

Another concern is spurious wave generation from the artificial core.
In our simulations, we observe sphere-shaped perturbations propagating from the core to the stellar surface.
The perturbations appear to be sound waves generated in the central regions of our simulations.
Given that the time interval of the episodic material ejection fits with the fundamental mode instead of the much shorter interval of these spurious waves, we think the spurious waves do not drive the material ejections, and therefore are dynamically not important.
However, it would be helpful to diminish the spurious waves, whose origins are not clear.
One possibility is that the convective velocities are damped too sharply in the artificial core that they introduce pressure perturbations on top of the hydrostatic equilibrium background.
Another possibility is that the hydrostatic equilibrium structure is modified due to different convective energy transport in 3D, which results in a small mismatch of entropy in the artificial core and the envelope (see Figure~\ref{fig:1d:1d3d}) and therefore seeds the pressure perturbations.

The spatial resolution of our simulations is not enough to resolve the photosphere.
This leads to a time-averaged luminosity output different from the input luminosity from the artificial core \citep[Figure~\ref{fig:0d:tvar}; for a test illustrating this, see Figure 17 in][]{ma2025Submitt.AA}.
We have performed short test simulations that suggest decreasing the cell size by a factor of two is still not enough to obtain the correct luminosity.
\citet{chiavassa2011A&Aa} also showed that increasing spatial resolution helps resolve smaller convective structures near the surface.
Another potential issue is whether the recombination layer inside the envelope is well-resolved in our simulations.
We have checked that our simulation resolution is enough to resolve the opacity peak due to He and H recombination, but it is not clear whether the variation in internal energy due to recombination is also resolved.
In 1D models, the recombination zones can be very thin during the contraction phase \citep{clayton2018, bronner2025arXive-prints}.
Further analyses are needed to assess if we resolve the recombination layer in 3D.

Despite using $80$ angles for radiation transport, we have found that fixing the discretized angle still results in ray-effects several stellar radii away from the star.
The ray-effects manifest as flower-like patterns with multiple peaks in radiation field in the optically-thin atmosphere \citep[e.g., Figure 16 in][]{ma2025Submitt.AA}.
Since the temperature is tightly coupled with radiation, and opacity depends strongly on temperature, the ray-effects also create artificial peaks in the spatial distribution of opacity.
This is particularly important for outflows driven by radiation pushing on dust, which depend critically on both the opacity and radiation field.
Such effects can in principle be avoided if we introduce extra diffusion by changing the transport directions intermittently \citep[e.g.,][]{freytag2012JCoPh, peter2023MNRAS}, which will be explored in the future.

Our treatment of the equation of state needs to be further improved.
For now, we transition from a radiation-included equation of state to a radiation-excluded one at density $10^{-9}$--$10^{-11}\, \mathrm{g\, cm^{-3}}$ (see Appendix~\ref{apx:method:rt} for details).
This means the radiation force is artificially reduced in the transition region near the stellar surface, which weakens the mass ejections.
In addition, the advection components of the radiation flux are neglected for density $<10^{-9}\, \mathrm{g\, cm^{-3}}$.
This is likely a reasonable approximation because the radiation flux is dominated by the comoving term in the cooling-dominated surface layer and atmosphere.
We also transition from the OPAL equation of state to an ideal gas at temperature $1870$ K.
Both of these `stitches' of the equation of state do not guarantee the consistency of thermodynamical variables.
Since the temperature field is also tightly coupled with radiation, we find that these `stitches' introduce artificially enhanced temperature at the transition boundaries in the RSG atmosphere.
These artificial temperature jumps have limited effects on the atmospheric dynamics, as they are confined within a thin layer in the transition boundary.
However, they likely need to be removed in post-processing for synthetic observations.

Finally, our simulations are limited to gray radiation hydrodynamics, and there are other important physics missing, e.g., a detailed treatment of dust, and magnetic fields.
Multi-group radiation transport or more realistic opacities are important for the atmospheric structure, in particular for the temperature stratification \citep[e.g.,][]{malygin2014A&A}.
How the dense molecular lines in the RSG atmosphere affect the dynamics is also not clear \citep{kee2021A&A}.
Another important physical ingredient is dust.
In this work, we only include the effects of dust as a high opacity below $1500$ K, but we ignore the detailed treatment of dust formation, advection, and momentum and energy exchange between radiation, dust, and gas.
In AGB star atmospheres, the dust growth timescale can be comparable to the pulsation period \citep{hofner2018A&AR}, which means a time-dependent treatment of dust grain growth is needed \citep{hofner2016A&A}.
One step forward would be to include part of those effects, as in e.g., \texttt{CO$^5$BOLD} simulations of AGB stars \citep{freytag2023A&A}.
We also do not include magnetic fields in our simulations so far.
The convective dynamo is expected to sustain a magnetic field.
Since the RSG surface pressure is dominated by turbulent pressure \citep{goldberg2022ApJ, chiavassa2024LRCA}, we also expect the magnetic field will be dynamically important, assuming the magnetic field energy is comparable to the convective energy.
Alfven wave dissipation may also provide another heating mechanism to launch the wind \citep[e.g.,][]{hartmann1980ApJ}.

\bibliography{RSG}{}
\bibliographystyle{aasjournal}

\end{CJK*}
\end{document}